\documentclass[compsoc,conference,a4paper,10pt,times]{IEEEtran}
\IEEEoverridecommandlockouts
\usepackage{amsmath}          
\usepackage{amssymb}          
\usepackage{amsthm}           
\usepackage{cite}
\usepackage{bm}
\usepackage{algorithm}
\usepackage{amsmath,amssymb,amsfonts}
\usepackage{algpseudocode} 
\usepackage{float}  
\usepackage{graphicx}
\usepackage{textcomp}
\usepackage{bmpsize}
\usepackage{lipsum}
\usepackage{multirow}
\usepackage{array}
\usepackage{bm}
\usepackage{graphicx} 
\usepackage{algorithm}
\usepackage{algpseudocode}
\usepackage{multirow}
\usepackage[most]{tcolorbox}
\usepackage{xcolor}
\usepackage{array}

\usepackage[colorlinks=true,urlcolor=black]{hyperref}
\def\BibTeX{{\rm B\kern-.05em{\sc i\kern-.025em b}\kern-.08em
    T\kern-.1667em\lower.7ex\hbox{E}\kern-.125emX}}
\usepackage{subcaption}

\newtheorem{theorem}{Theorem}
\newtheorem{definition}{Definition}

\newcommand{\negl}{\nu}

\newcommand{\Exp}{\textsc{Exp}}

\begin{document}

\title{ZKPROV: A Zero-Knowledge Approach to Dataset Provenance for Large Language Models}

\author{
\IEEEauthorblockN{Mina Namazi}
\IEEEauthorblockA{\textit{Open University of Catalonia} \\
Barcelona, Spain \\
mnamaziesfanjani@uoc.edu}
\and
\IEEEauthorblockN{Alexander Nemecek}
\IEEEauthorblockA{\textit{Case Western Reserve University}\\
Cleveland, OH, USA \\
ajn98@case.edu}
\and
\IEEEauthorblockN{Erman Ayday}
\IEEEauthorblockA{\textit{Case Western Reserve University}\\
Cleveland, OH, USA \\
exa208@case.edu}
}

\maketitle

\begin{abstract}
As large language models (LLMs) are used in sensitive fields, accurately verifying their computational provenance without disclosing their training datasets poses a significant challenge, particularly in regulated sectors such as healthcare, which have strict requirements for dataset use. Traditional approaches either incur substantial computational cost to fully verify the entire training process or leak unauthorized information to the verifier. Therefore, we introduce ZKPROV, a novel cryptographic framework allowing users to verify that the LLM's responses to their prompts are trained on datasets certified by the authorities that own them. Additionally, it ensures that the dataset's content is relevant to the users' queries without revealing sensitive information about the datasets or the model parameters. ZKPROV offers a unique balance between privacy and efficiency by binding training datasets, model parameters, and responses, while also attaching zero-knowledge proofs to the responses generated by the LLM to validate these claims. Our experimental results demonstrate sublinear scaling for generating and verifying these proofs, with end-to-end overhead under $3.3$ seconds for models up to $8$B parameters, presenting a practical solution for real-world applications. We also provide formal security guarantees, proving that our approach preserves dataset confidentiality while ensuring trustworthy dataset provenance.
\end{abstract}


\section{Introduction}
\label{intro}
Large language models (LLMs) are deployed in today's critical decision-making processes, including healthcare diagnostics~\cite{bhavsar2021medical}, financial risk assessment~\cite{addo2018credit}, and legal services~\cite{armour2020ai}; however, they introduce significant challenges in verifying their computational integrity~\cite{aswathy2025machine, li2023trustworthy}, particularly in regulated domains where reliability is legally required. Current LLM deployments lack efficient mechanisms to prove that models were trained on authorized, relevant datasets without revealing the individual's sensitive information stored in the training datasets, creating a critical gap between regulatory compliance requirements (such as GDPR~\cite{regulationgeneral} or HIPAA~\cite{hipaa1996}) and technical capabilities. 

Verification methods in the LLM field are categorized into three core areas: (i) inference verification to validate model outputs, (ii) training process verification to confirm proper algorithm execution, and (iii) training data verification to ensure models are trained on authorized datasets. Malicious actors may compromise these aspects to reduce computational costs or introduce privacy vulnerabilities, particularly when computations are outsourced due to privacy concerns or resource limitations~\cite{xing2023zero}. 

Several cryptographic approaches have emerged to address different aspects of machine learning integrity. Secure Multi-party Computation (SMC) enables confidential, collaborative training but lacks scalability and requires constant connectivity among participants~\cite{zhang2020zkdt_zkpytorch_ref26, kang2023scaling_zkpytorch_ref15}. Trusted Execution Environments (TEEs) offer hardware-based integrity guarantees but remain vulnerable to side-channel attacks and rely on centralized trust assumptions~\cite{tramer2018slalom}. Homomorphic Encryption (HE) allows computation on encrypted data, but introduces performance overhead for complex models~\cite{gilad2016cryptonets}. 

Zero-knowledge proofs (ZKPs) demonstrate a promising path for verifiable machine learning. For example, systems such as VeriML~\cite{zhao2021veriml}, zkCNN~\cite{liu2021zkcnn_zkpytorch_ref16}, zkLLM~\cite{sun2024zkllm}, and zkGPT~\cite{lu2024efficient} focus on verifying the training process or inference correctness, but not on validating training data provenance, the ability to cryptographically prove that a model was trained on a precise, authorized datasets without revealing information about their content or the LLM model's parameters.

This paper addresses the gap in current zero-knowledge-based approaches to LLM verification. While current approaches~\cite{chen2024zkml,feng2021zen,feng2024zeno_zkpytorch_ref9,hao2024scalable,kang2022scaling,liu2021zkcnn_zkpytorch_ref16,lu2024efficient,sun2024zkllm,zhang2020zero,maheri2025telesparse} can prove the computational correctness of training or inference, they cannot cryptographically verify that a model was trained on a specific, authorized dataset without revealing the dataset itself or the model architecture. This feature is essential in settings where regulatory compliance requires models to draw only from relevant data sources, including credible information about the corresponding prompts. For example, if a medical professional queries an LLM with ``What is the life expectancy of patients diagnosed with breast cancer in Georgia hospitals between 1985 and 2025?'', ZKPROV proves that the model generating this response was trained on authenticated clinical datasets from Georgia hospitals that include breast cancer patient data within that time period. Our evaluation demonstrates that ZKPROV achieves this verification with practical overhead, generating responses with cryptographic provenance proofs that require under $3.3$ seconds end-to-end for models up to $8$B parameters, with proof generation and verification each completing in under $1.8$ seconds.
 
This work introduces a novel framework, ZKPROV, to verify dataset provenance without the computational burden of proving the entire training process. Our approach generates verifiable proofs that cryptographically bind a model's response to its training dataset while preserving the privacy of both the dataset and the model parameters. Our main \textit{contributions} are listed as follows.

\begin{itemize}
 \item We design a novel, privacy-preserving framework that cryptographically binds LLM responses to relevant, authorized training datasets without revealing dataset contents or model internals.
 
\item We introduce unique binding values to link the LLM's phases of the query-response process and eliminate the verification effort required for the training process's computation steps.

\item We deploy efficient signature schemes, commitments, and recursive zero-knowledge proofs to verify LLM's data provenance without revealing the datasets' content.

\item We achieve sublinear scaling for generating proofs and verifying dataset relevance across multiple transformer layers, with experimental results on Llama models ($1$B-$8$B parameters) demonstrating consistent cryptographic overhead of $1.8$ seconds or less regardless of model size.
 
 \item We formally prove that the proposed framework is robust and preserves the privacy of the training dataset and model parameters in LLM.
\end{itemize}

The remainder of this paper is organized as follows: Section~\ref{related} reviews related work in verifiable machine learning and zero-knowledge proofs. Section~\ref{back} provides the background required to design our framework in Section~\ref{proposed}. Section~\ref{sec} analyses the security and privacy of our proposed protocol. Section~\ref{eval} presents our experimental evaluation. Section~\ref{discus} discusses limitations and future work, and Section~\ref{conc} concludes the paper.

\section{Related Work}
\label{related}

This section reviews related work in verifiable machine learning, highlighting the deployment of zero-knowledge proofs (ZKPs).

Feng et al. introduced ZEN~\cite{feng2021zen}, which provides inference verification for large neural networks by quantizing and encoding computations to reduce proof cost. They focus on proving that the inference computations are correct with respect to committed model parameters and achieve verification times ranging from seconds to minutes for large networks. Verification has also been extended to LLMs, such as in zkLLM~\cite{sun2024zkllm}, which optimizes the arithmetization of non-linear operations such as attention mechanisms and activation functions. The zkLLM introduces specialized components to handle transformer operations at scale. These techniques allow a prover to demonstrate that a specific input-output pair is produced by a committed model, without exposing internal model details. Inference verifications are developed for other model classes, such as convolutional neural networks~\cite{liu2021zkcnn_zkpytorch_ref16} and decision trees~\cite{zhang2020zero}, using circuit optimizations with respect to the model's specifications.

These solutions' key limitations are verifying inference correctness after training the model, assuming that all training data is authorized. However, they offer no assurance regarding which datasets are used during training. The limitations are more challenging in domains that require dataset auditability and regulatory compliance, such as healthcare, where outputs must be derived exclusively from institutionally approved datasets that contain relevant information for the user query.  

Beyond verifying model outputs, methods are suggested to verify the training process using ZKPs. These protocols prove that a model is trained correctly, commonly via gradient descent, on a committed dataset~\cite{abbaszadeh2024zero, garg2023experimenting}. Such approaches provide verifiability for the entire computational training process by requiring proofs for each optimization step. However, as model sizes and training complexity grow, these methods become unscalable.

Moreover, in these systems, no explicit cryptographic guarantees are provided regarding the authenticity of training data from approved sources. Recent work highlights that data authenticity, consent, and provenance remain fundamental challenges for AI systems, with no complete solution existing for verifiable dataset provenance~\cite{longpre2024data}. While authentication and provenance mechanisms have been proposed to prevent data poisoning attacks in machine learning, these focus primarily on detecting malicious alterations rather than verifying authorized dataset usage~\cite{stokes2021preventing}. Even if the training procedure is proven to follow a prescribed algorithm, there is no mechanism to bind the resulting model to specific, authorized datasets.

Contrasting our approach with retrieval-augmented generation (RAG), which retrieves relevant information from a hosted dataset at inference time, is essential. While RAG systems have emerged as a prominent approach to enhance LLM outputs by incorporating external knowledge~\cite{gao2023retrieval, gupta2024comprehensive}, they face challenges with trustworthiness, factuality, and accountability~\cite{zhou2024trustworthiness}. Although RAG implementations can restrict information retrieval through authorization levels~\cite{aws2024rag, aws2024ragauth}, they do not verify that the underlying LLM was trained exclusively on authorized datasets. In regulated domains like finance and healthcare, where compliance with data protection regulations is critical~\cite{fitzgerald2024rag}, it is necessary to ensure that models themselves are constrained by authorized training datasets, not just retrieval sources. Our framework addresses this need by enabling cryptographic verification that a model response originated from a model bound to such datasets.

Although RAG techniques could potentially be integrated into the ZKPROV framework, our current focus is on verifying dataset provenance for trained models, independent of how the training is performed. While our evaluation fine-tunes models to demonstrate domain relevance, the ZKPROV protocol determines whether the model is pre-trained, fine-tuned, or trained from scratch. This choice reflects a practical deployment model where training and serving a model with verified dataset provenance is simpler than securely hosting and governing access to the datasets themselves at inference time. We further discuss this integration opportunity in Section~\ref{discus}.

We introduce a new category in the field of verifiable machine learning with privacy-preserving dataset provenance for LLMs. Unlike prior work, our framework requires no verification of the complete training process. Instead, it provides ZKPs that a model is trained or fine-tuned using relevant, authenticated datasets, while keeping the contents confidential.

\section{Background}
\label{back}
We deploy state-of-the-art cryptographic schemes that efficiently generate and verify proofs to validate the authenticity and relevance of the training datasets while responding to the users' queries. We introduce the building blocks and encryption schemes required to design our proposed privacy-preserving framework for dataset-provenance verification in large language models (LLMs). 

\subsection{Large Language Model Primitives}
\label{LLM}

Given the sensitivity of its potential applications, such as healthcare, we must ensure that the final model response $r$ is provably linked to a specific, authorized dataset.

We define the fine-tuning process for the initial LLM model weights $W_0$ and dataset $D$ as: 
$$W \leftarrow \textsc{FineTune}(D, W_0, H).$$
Where $H = (\eta, B, E, O)$ specifies the learning rate, batch size, epochs, and optimizer, respectively, to define the acceptable training configurations for regulatory compliance.




\subsection{Cryptographic Primitives}
\label{primit}

We describe the core cryptographic schemes and their hardness assumptions, which are essential for developing the proposed privacy-preserving LLM provenance verification framework. 

 Let $\mathbb{G}_1$ and $\mathbb{G}_2$ be elliptic curve groups of prime order $q$, and let $\mathbb{G}_T$ denote the target group. A bilinear map $e: \mathbb{G}_1 \times \mathbb{G}_2 \rightarrow \mathbb{G}_T$ satisfies:
\begin{itemize}

\item  Bilinearity: For all $g_1 \in \mathbb{G}_1$, $g_2 \in \mathbb{G}_2$, and $a, b \in \mathbb{Z}_q$, the following equation holds, $e(g_1^a, g_2^b) = e(g_1, g_2)^{ab}$.
    
\item  Non-degeneracy: $e(g_1, g_2) \neq 1$, where $g_1$ and $g_2$ are generators of $\mathbb{G}_1$ and $\mathbb{G}_2$, respectively.
\end{itemize}

\subsubsection{Zero-Knowledge Succinct Non-Interactive Argument of Knowledge (zk-SNARK)}
\label{zkp}
A zk-SNARK is a zero-knowledge proof system that allows a prover to convince a verifier that a statement \( x \in \mathcal{L}_R \) is valid with respect to a relation \( R \), without revealing any information beyond the statement. It compiles the required computations as constraint systems (typically Rank-1 Constraint Systems (R1CS)). A constraint system defines a relation $R$ such that a statement $x$ is valid if and only if there exists a witness $\omega$ satisfying the constraints encoded in $R$. 

In our proposed framework for verifying the provenance of LLMs, computations across each transformer layer must be efficiently verified. Therefore, we use HyperNova~\cite{kothapalli2024hypernova} to leverage its folding mechanism and recursively fold the generated proofs for correct computations of each layer into a single succinct one. The folding mechanism allows complex LLM computations involving high-degree polynomial operations. It consists of three main components as follows.

HyperNova comproses $\text{HN} = \{\textsc{ZK.Setup}, \textsc{ZK.Prove}, \textsc{ZK.Verify}, \textsc{ZK.Fold}\}$, described as follows.

\begin{itemize}
    \item $\textsc{ZK.Setup}(1^\lambda, R) \rightarrow pp_{zk}$: Given a security parameter $\lambda$ and a relation $R$, the algorithm generates common reference strings (crs) for the underlying polynomial commitment scheme, encodes the constraint matrices as polynomials, and outputs $pp_{zk} = \{\text{crs}, pk, vk\}$, where $pk$ is the proving key and $vk$ is the verification key. The $pp_{zk}$ is part of the entire algorithm inputs.
    
    \item $\textsc{ZK.Prove}(pk, x, \omega) \rightarrow \pi$: Inputs the proving key $pk$, a public statement $x$, and a private witness $\omega$, and generate a proof $\pi$ for $x$.
    
    \item $\textsc{ZK.Verify}(vk, x, \pi) \rightarrow \{0,1\}$: Takes the verification key $vk$, the public input $x$, and the proof $\pi$. It validates the proof through checking, $e(C, g) = e(\pi, g_{\text{crs}})$, where $C = g^{f(\alpha)}$ is the commitment to the polynomial $f(X)$.
    
    \item $\textsc{ZK.Fold}(pk, (U_1, w_1), (U_2, w_2)) \rightarrow (U', w')$: The algorithm combines two instances $(U_1, w_1)$ and $(U_2, w_2)$ into a single one $(U', w')$. It compresses multiple computational steps using a random linear combination.
\end{itemize}

 HyperNova's verification complexity is logarithmic in the size of the constraint system, enabling efficient verification for LLM computations. 
In the following, we present the security properties of this proof system.

{\textbf{Negligible Function.}}
A function $\negl: \mathbb{N} \rightarrow \mathbb{R}_{\ge 0}$ is negligible if for every positive polynomial $p(\cdot)$, there exists an $N_p$ such that for all $n > N_p$, $\negl(n) < \frac{1}{p(n)}$.

{\textbf{Computational Soundness.}} 
A ZKP  satisfies computational soundness if, for every probabilistic polynomial-time (PPT) adversary \(Adv\):
\[
\Pr\left[\textsc{ZK.VERIFY}(vk, x, \pi) = 1 \, \land \, x \notin \mathcal{L}_R \right] \leq \negl(\lambda).
\]

{\textbf{Zero-Knowledge.}} 
A ZKP protocol is zero-knowledge if for every PPT verifier \(\mathcal{V}^*\), there exists a PPT simulator \(\mathcal{S}\) such that for all \(x \in \mathcal{L}_R\) with witness \(w\), the view of \(\mathcal{V}^*\) in a real interaction with the prover is computationally indistinguishable from the output of \(\mathcal{S}\) that does not have access to \(w\):
\[
\left\{\text{View}_{\mathcal{V}^*}(\textsc{ZK.Prove}(pk, x, w))\right\} \approx_c \left\{\mathcal{S}(pk, x)\right\}.
\]
Where \(\text{View}_{\mathcal{V}^*}\) denotes the view of \(\mathcal{V}^*\) during the protocol, and \(\approx_c\) denotes computational indistinguishability.




\subsubsection{Commitment Scheme}
\label{commitscheme}

A commitment scheme is a cryptographic primitive that allows a party to commit to a value while keeping it hidden. They are usually hiding; no information is revealed about the committed value. They are also binding, meaning that once a commitment is created, it is computationally infeasible to change it to a different value. The message and opening information can be revealed later.

The Hypernova scheme internally deploys the Kate-Zaverucha-Goldberg (KZG)~\cite{kate2010constant} to commit to polynomials \( f(X) \). It later proves the evaluation of the polynomial at any point \( z \) without revealing the entire polynomial. It consists of KZG = \{\textsc{C.SETUP}, \textsc{C.COMMIT}, \textsc{C.OPEN}, and \textsc{C.VERIFY}\}, as follows.
\begin{itemize}
    \item $\textsc{C.Setup}(1^\lambda, d) \rightarrow pp_c$: Samples $\alpha \xleftarrow{\$} \mathbb{F}_p$, generates public parameters $pp_{c} = (g, g^\alpha, g^{\alpha^2}, \ldots, g^{\alpha^d}) \in \mathbb{G}_1^{d+1}$. These parameters are part of the all algorithms in the protocol.
    
    \item $\textsc{C.Commit}(f(X), \alpha) \rightarrow C$: For polynomial $f(X) = \sum_{i=0}^{d} f_i X^i$, it generates commitment $C = \prod_i (g^{\alpha^i})^{f_i}$. 
    where $f_i \in \mathbb{F}_p$ are the coefficients and $d$ is the degree bound, and $\alpha$ is a secret value.
    \item $\textsc{C.Open}(f(X), z) \rightarrow \pi$: Generates an opening proof $\pi$ for evaluation $f(z)$ at point $z$:
    \begin{equation*}
    \pi = g^{q(\alpha)} \text{ where } q(X) = \frac{f(X) - f(z)}{X - z}.
    \end{equation*}
    The proof $\pi$ ensures that the prover knows the entire polynomial $f(X)$ without revealing it. The quotient polynomial $q(X)$ guarantees the correctness of $f(z)$.
    
    \item $\textsc{C.Verify}(C, z, y, \pi) \rightarrow \{0,1\}$: Checks correctness of $f(z) = y$ using pairing:
    \begin{equation*}
    e(C \cdot g^{-y}, g) = e(\pi, g^{\alpha} \cdot g^{-z}).
    \end{equation*}
    This pairing equation ensures that the committed polynomial evaluates correctly at $z$, and without the verifier knowing $\alpha$ or $f(X)$.
\end{itemize}


\subsubsection{Tree-Based Commitment Scheme}
\label{tree}
 We deploy Reckle Trees~\cite{papamanthou2024reckle} to organize the elements of the datasets, given their efficiency in supporting batch updates. Particularly, in healthcare scenarios, where datasets frequently require new patient records, updated treatment protocols, or regulatory compliance updates, proving dataset membership for multiple data points simultaneously becomes highly efficient. 

 \subsubsection{Boneh-Lynn-Shacham}
 \label{BLS}
 We use Boneh-Lynn-Shacham (BLS)~\cite{boneh2001short} signature scheme to authenticate the data. The BLS provides short signatures with efficient verification, providing strong unforgeability under chosen-message attacks. It enables signature aggregation for batch verification, which is critical for minimizing communication overhead and verifying multiple datasets signed by different authorities simultaneously. Its algorithms are described as follows.
 \begin{itemize}

 \item \textsc{BLS.Sign}$(sk, m) \rightarrow \sigma$:  
Takes the secret key $sk$ and message $m$, computes the signature $\sigma$.

\item \textsc{BLS.Verify}$(pk, m, \sigma) \rightarrow \{0,1\}$:  
Takes public key $pk$, message $m$, and the signature $\sigma$, and outputs $1$ if the signature is valid, $0$ otherwise.
 \end{itemize}

\section{Proposed Scheme}
\label{proposed}
\begin{figure*}[t]
    \centering
    \includegraphics[width=\textwidth]{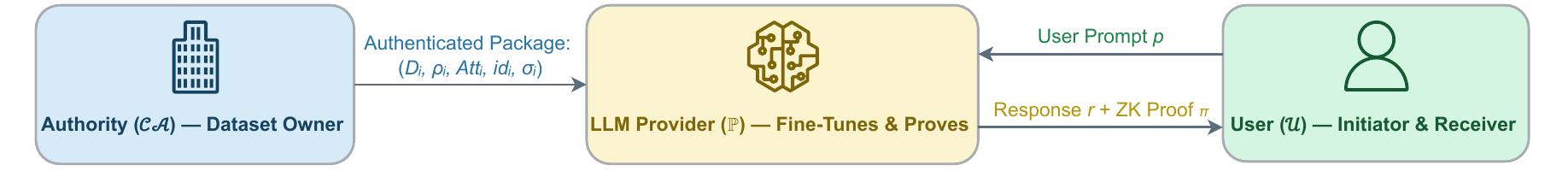}
\caption{High-level ZKPROV framework's flow. The figure illustrates the interaction between $\mathcal{CA}$, $\mathcal{P}$, and $\mathcal{U}$. $\mathcal{CA}$ certifies dataset authenticity and transmits signed metadata to $\mathcal{P}$, who fine-tunes a base model on the authenticated datasets and produces provenance-aware responses. $\mathcal{U}$ submits a prompt with explicit attributes and receives a response along with a zero-knowledge proof attesting that the model is trained on authenticated datasets, including a dataset with attribute alignment. The labeled arrows highlight the authenticated dataset handoff, prompt input, and proof-backed output.}
\label{Fig:overview}
\end{figure*}

This section introduces ZKPROV, a robust privacy-preserving framework for training data provenance in LLMs. To the best of our knowledge, it is the first system to bridge the gap between data regulation and LLM auditability. It enables users to verify that the LLM responds by using a model trained on approved datasets that contain relevant information for the user prompt, without verifying the entire training or inference process. At the same time, no information about the model, dataset, or training process is leaked to the user during verification. We outline the proposed scheme's setting, threat model, overview, and detailed description here, and provide an overview in Figure~\ref{Fig:overview}.

\subsection{System Model and Setting}
\label{systemmodel}

The proposed ZKPROV comprises an authority $\mathcal{CA}$ (e.g., healthcare institution) that owns and authenticates datasets, a prover $\mathcal{P}$ (the LLM service provider) that operates a language model, and a user $\mathcal{U}$ who submits queries and verifies proofs.

 The authority $\mathcal{CA}$ owns a collection of datasets $\mathcal{D} = \{D_1, \ldots, D_m\}$ where each dataset $D_i$ is characterized by attribute sets $Att_i = \{att_1, \ldots, att_k\}$ reflecting demographic information (e.g., $Att_1 = \{\text{``breast\_cancer''}, \text{``female''}, \text{``age\_65+''}\}$) and includes administrative identifiers $\text{Id}_i = \{id_1, \ldots, id_k\}$ for dataset provenance tracking. Each dataset $D_i$ is organized in a Reckle Tree $T_i$ with root $\rho_i$, which serves as a commitment to the dataset's contents. 

The prover $\mathcal{P}$ operates a model $W$ with $N$ layers trained on base model $W_0$ using the authenticated dataset collection $\mathcal{D}$. The authenticated datasets used for training are represented by their roots $\{\rho_1, \ldots, \rho_m\}$, which are organized in a training set tree $T_{tr}$ with root $R_{tr}$. For each layer $j \in \{1, \ldots, N\}$, the weight differences $\Delta W_j = W_j - W_{0,j}$ capture the parameter transformations induced by training on the authenticated datasets.

 A query consists of a pair $(p, Att_p)$ where $p$ is natural language text and $Att_p$ is an attribute set characterizing domain requirements (e.g., $Att_p = \{\text{``breast\_cancer''}, \text{``prognosis''}, \text{``elderly''}\}$). A response $r$ consists of natural language text accompanied by a proof $\pi$ that verifiably demonstrates the model was trained on authenticated datasets, including a dataset with relevant attributes to the query domain.

\subsection{Threat Model}
\label{threat}

The authority $\mathcal{CA}$ is assumed to be honest, to own the datasets, to construct Reckle Trees with correct roots, to form metadata packages with correct attributes, and to generate signatures. 

The user $\mathcal{U}$ is semi-honest, meaning they follow the protocol instructions correctly but may be curious about unauthorized information about the dataset contents or model parameters. 

The LLM service provider $\mathcal{P}$ may be malicious and attempt to use unauthorized data, forge protocol outputs, or access the datasets' content. A malicious $\mathcal{P}$ may attempt to train models on unauthorized or modified datasets while claiming to use authenticated ones, or to reuse proofs across different query contexts to decouple provenance verification from the actual response generation.

Formally, in the ZKPROV framework, the following security properties are guaranteed.
\begin{itemize}
    \item \textbf{Zero-Knowledge (ZK)}: The proofs of provenance must not reveal any information about the model weights, or the underlying training datasets, beyond what is explicitly disclosed through the public commitments to dataset metadata and the published training set root.
    
    \item \textbf{Soundness}: A malicious prover cannot convince an honest verifier of a false statement, meaning they cannot pass a proof generated on unauthorized or irrelevant datasets
    
   \item \textbf{Transcript Binding}: A proof generated for a specific query-response pair must not be valid for any different pair. This prevents a malicious prover from pre-computing proofs before receiving queries or reusing proofs across different interactions. 
\end{itemize}

\subsection{Overview}
\label{overview}

 $\mathcal{CA}$ stores the data in a Reckle Tree structure obtaining root $\rho_i$, constructs metadata $m_i = (\rho_i, Att_i, id_i)$, and generates BLS signature $\sigma_i$ on the metadata using secret key $sk_{\mathcal{CA}}$, creating unforgeable authentication tokens, and transmits the authenticated datasets along with their signatures and metadata to $\mathcal{P}$. 

$\mathcal{P}$ uses these authenticated datasets to fine-tune a base model $W_0$, obtaining a model $W$, and forms a tree of training sets with roots stored in $R_{tr}$. Then it commits to signatures $\{\sigma_i\}_{i=1}^m$, metadata $\{m_i\}_{i=1}^m$, weight differences $\{\Delta W_j\}_{j=1}^N$. Then they bound these values together with $R_{tr}$, and the base model $W_0$.

 At query time, when $\mathcal{U}$ submits query $(p, Att_p)$ consisting of a prompt and the attributes, $\mathcal{P}$ generates response $r$ using the trained model. $\mathcal{P}$ constructs a proof $\pi$ demonstrating that the model was trained on authenticated datasets relevant to the query. The proof construction derives challenge vectors for each layer from the fixed commitments, the root of the training dataset, and the query, ensuring that the challenges cannot be manipulated. The proof links the response, the prompt, the query attributes, and the relevant dataset while preserving the datasets' privacy.

  $\mathcal{U}$ receives all the proofs and verifies that the datasets are authenticated and members of the training collection. Also, the weight differences are bound to the training set, binding values are computed correctly for all layers, and the proof corresponds to this specific query-response interaction.

Our ZKPROV's main algorithms are defined as follows.

\noindent ZKPROV = $\{\textsc{Setup}, \textsc{Prove}, \textsc{Verify}\}$ as defined in the following.

\begin{itemize}
\item $\textsc{Setup}(1^\lambda, \{D_i, Att_i\}_{i=1}^m, W_0) \rightarrow (\mathcal{C}, \Omega)$: Given security parameter $\lambda$, datasets with attributes $\{D_i, Att_i\}_{i=1}^m$, and the base model $W_0$, it generates the necessary cryptographic parameters for all participants. performs dataset authentication and model fine-tuning on the authenticated datasets, and produces public commitments $\mathcal{C}$ and private witnesses $\Omega$ for all sensitive protocol components.

\item $\textsc{Prove}(\mathcal{C}, \Omega, p, Att_p) \rightarrow (r, \pi)$: Given commitments $\mathcal{C}$, witnesses $\Omega$, prompt $p$, and query attributes $Att_p$, generates response $r$ using the trained model and produces a proof $\pi$ demonstrating that the model was trained on authenticated datasets including a dataset with relevant attributes to the user query.

\item $\textsc{Verify}(p, Att_p, r, \pi, \mathcal{C}) \rightarrow \{\textsc{Accept}, \textsc{Reject}\}$: Given prompt $p$, query attributes $Att_p$, response $r$, proof $\pi$, and commitments $\mathcal{C}$, validates all cryptographic relationships to confirm that the model was trained on authenticated datasets including a dataset with attribute alignment without learning sensitive information about dataset contents or model parameters.
\end{itemize}

\subsection{Detailed Description of ZKPROV}
\label{detail}

The ZKPROV framework comprises three algorithms for establishing and verifying dataset provenance for LLM responses. We use Reckle trees~\cite{papamanthou2024reckle} to succinctly commit to dataset contents, BLS signatures~\cite{boneh2001short} to authenticate datasets and bind them to their attributes, binding formulas to establish statistical connections between model weights and the training set, and HyperNova~\cite{kothapalli2024hypernova} to verify the dataset's relevance to the generated response efficiently.

The three phases of ZKPROV as defined in Sec.~\ref{overview} are described in detail as follows.

\subsubsection{Setup and Preprocessing}
\label{phase1}

The setup phase implements the function:
\begin{equation*}
\textsc{Setup}(1^\lambda, \{D_i, Att_i\}_{i=1}^m, W_0) \rightarrow (\mathcal{C}, \Omega).
\end{equation*}
Where $\lambda$ is the security parameter, $\{D_i, Att_i\}_{i=1}^m$ are the datasets with their attributes, $W_0$ is the base model, $\mathcal{C}$ denotes the public commitments, and $\Omega$ denotes the private witnesses.

$\mathcal{CA}$ generates the common reference string $\text{crs} = (\kappa_1, \kappa_2, \kappa_3, \kappa_4)$ where $\kappa_i$'s  are the domain separators for various parts of the protocol. 
$\mathcal{CA}$ generates a BLS key pair $(sk_{\mathcal{CA}}, pk_{\mathcal{CA}}) \gets \textsc{BLS.KeyGen}(1^\lambda)$ according to the key generation procedure described in Sec.~\ref{BLS}.

For each dataset $D_i$ where $i \in \{1, \ldots, m\}$, $\mathcal{CA}$ performs dataset authentication. It constructs a Reckle Tree $T_i$ for dataset $D_i$ (Sec.~\ref{tree}), obtaining root $\rho_i$. Then it forms metadata $m_i = (\rho_i, Att_i, id_i)$ that combines the tree root, the attributes $Att_i \in \mathcal{A}$ showing the domain of the datasets, and an identifier $id_i$ for dataset tracking. $\mathcal{CA}$ signs the metadata using $\sigma_i \gets \textsc{BLS.Sign}(sk_{\mathcal{CA}}, m_i)$, binding the dataset to its attributes, and transmits the authentication package $AP = (pp, \{D_i, m_i, \sigma_i, T_i\}_{i=1}^m)$ to $\mathcal{P}$ where $pp = (\text{crs}, pk_{\mathcal{CA}})$ are the public parameters entering to the rest of the defined algorithms.

Upon receiving $AP$, for each dataset $i \in \{1, \ldots, m\}$, $\mathcal{P}$ verifies the signature by calling $\textsc{BLS.Verify}(pk_{\mathcal{CA}}, m_i, \sigma_i) \stackrel{?}{=} 1$. If verification fails, the protocol aborts. Upon successful verification of the signatures, $\mathcal{P}$ starts fine-tuning various models based on the authenticated training sets by calling $W \gets \textsc{FineTune}(\{D_i\}_{i \in S}, W_0, H)$, where $S \subseteq \{1, \ldots, m\}$ is the set of dataset indices selected for training, with hyperparameters $H = (\eta, B, E, O)$ representing learning rate, batch size, number of epochs, and optimizer configuration respectively. For each layer $j \in \{1, \ldots, N\}$, $\mathcal{P}$ computes the weight difference $\Delta W_j = W_j - W_{0,j}$ where $W_j, W_{0,j} \in \mathbb{R}^{d_j}$ denote the weights of layer $j$ in the fine-tuned and base models, capturing the specific parameter transformations induced by training on the authenticated datasets.

$\mathcal{P}$ constructs a tree of the training dataset with roots $R_{tr}$, and generates commitments to these components (Sec.~\ref{commitscheme}). For the base model, it samples uniform randomness $\omega_{W_0} \xleftarrow{\$} \mathbb{F}_p$ and computes commitment $C_{W_0} \gets \textsc{C.Commit}(W_0, \omega_{W_0})$. For the weight differences, it samples $\omega_{\Delta W} \xleftarrow{\$} \mathbb{F}_p$ and computes $C_{\Delta W} \gets \textsc{C.Commit}((\{\Delta W_j\}_{j=1}^N, R_{tr}), \omega_{\Delta W})$, binding the weight differences to the training set root within the commitment. For each dataset $i \in S$, $\mathcal{P}$ samples independent randomness values $\omega_{\sigma,i}, \omega_{m,i} \xleftarrow{\$} \mathbb{F}_p$, where the independence ensures cryptographic separation between different protocol components and prevents correlated input attacks. $\mathcal{P}$ computes commitments $C_{\sigma,i} \gets \textsc{C.Commit}(\sigma_i, \omega_{\sigma,i})$ for the signature and $C_{m,i} \gets \textsc{C.Commit}(m_i, \omega_{m,i})$ for the metadata. We ensure that these commitments are generated during the setup phase and become public before any user query arrives. 

$\mathcal{P}$ assembles the public commitments $\mathcal{C} = \{C_{m,i}, C_{\sigma,i}\}_{i=1}^m \cup \{C_{\Delta W}, C_{W_0}, R_{tr}\}$ and stores the private witnesses $\Omega = \{m_i, \omega_{m,i}, \sigma_i, \omega_{\sigma,i}\}_{i=1}^m \cup \{\{\Delta W_j\}_{j=1}^N, \omega_{\Delta W}, W_0, \omega_{W_0}, T_{tr}\}$ which will be used during proof generation to demonstrate knowledge of the committed values and prove various relationships between them.

The process of challenge vector derivation ensures different layers receive distinct challenges. $\kappa_2$ is the domain separator for the seed computation, preventing collisions. $\kappa_3$ is the domain separator for query transcript binding, ensuring transcripts are separated from challenge derivations.

\subsubsection{Proof Construction}
\label{proofconstruct}
The proof construction phase implements the function:
\begin{equation*}
\textsc{Prove}(\mathcal{C}, \Omega, p, Att_p) \rightarrow (r, \pi).
\end{equation*}
Where $\mathcal{C}$ contains the public commitments from setup, $\Omega$ contains the private witnesses, $p$ is the user's prompt, $Att_p$ are the query attributes, $r$ is the generated response, and $\pi$ is the zero-knowledge proof.

The user $\mathcal{U}$ submits query $(p, Att_p)$ and the LLM provider $\mathcal{P}$ generates the response $r \gets \textsc{LLM}(W, p)$ and constructs proofs establishing the provenance chain.

Let $\omega_\sigma = (\sigma_i, m_i, \omega_{\sigma,i}, \omega_{m,i})$ denote the signature witness. To prove that the datasets with index $i \in R_{tr}$, used for training the model, are authenticated and $\sigma_i$ is a valid signature on metadata $m_i$ under $\mathcal{CA}$'s public key,  $\mathcal{P}$ generates:
\begin{align*}
\pi_\sigma \gets \textsc{ZK.Prove}(pp_{zk}, (C_{m,i}, C_{\sigma,i}, pk_{\mathcal{CA}}), w_1).
\end{align*}
Where $w_1$ contains the signature $\sigma_i$ and the metadata $m_i$.

To prove that the training dataset with root $\rho_i$ extracted from committed metadata $m_i$ is a member of the training set it generates:
\begin{align}
\pi_{tr} \gets \textsc{ZK.Prove}(pp_{zk}, &(C_{\Delta W}, C_{m,i}), \nonumber
\omega_{2}).
\end{align}
Where $\omega_2=(R_{tr}, \rho_i, m_i, \omega_{\Delta W}, \omega_{m,i})$ is the private witness including the training set root $R_{tr}$, dataset root $\rho_i$, and metadata $m_i$.

To prove that $C_{\Delta W}$ (generated before the query) correctly reflects that the weight differences are bound to the same training set root, binding model parameters and training data, $\mathcal{P}$ generates:
\begin{align}
\pi_{bind} \gets \textsc{ZK.Prove}(pp_{zk}, &(C_{\Delta W}), \nonumber
\omega_{3}).
\end{align}

Where $\omega_{3} = (\{\Delta W_j\}_{j=1}^N, R_{tr}, \omega_{\Delta W})$

After generating the commitments and proofs and the query available now, $\mathcal{P}$ computes the challenge seed that transforms the protocol into a non-interactive form. Since all commitments were fixed during setup before the query arrived, and the seed includes the query and response, a malicious prover cannot manipulate the seed to produce favorable challenges. 
\begin{align}
seed \gets \textsc{Hash}(&C_{m,i} \parallel C_{\sigma,i} \parallel C_{\Delta W} \parallel \nonumber\\
&C_{W_0} \parallel Att_p \parallel p \parallel r \parallel \kappa_2)
\end{align}
For each layer $j \in \{1, \ldots, N\}$, $\mathcal{P}$ derives challenge vector $v_j \gets \textsc{Hash}(seed \parallel j \parallel \kappa_1)$ and computes binding value $B_j \gets \langle \Delta W_j, v_j \rangle$. $\mathcal{P}$ then samples $\omega_B \xleftarrow{\$} \mathbb{F}_p$ and commits $C_B \gets \textsc{C.Commit}(\{B_j\}_{j=1}^N, \omega_B)$.

Using HyperNova's recursive folding~\cite{kothapalli2024hypernova}, $\mathcal{P}$ folds all $N$ binding constraints into a single accumulator. Starting with $S_{B,0} \gets \emptyset$, for each layer $j$, it computes $S_{B,j} \gets \textsc{HN.Fold}(S_{B,j-1}, \cdot, \cdot)$. After processing all layers, $\mathcal{P}$ generates:
\begin{align}
\pi_B^{rec} \gets \textsc{HN.Prove}(pp_{zk}, S_{B,N}, \omega_4)
\end{align}

To prove dataset relevance, demonstrating that the query attributes are contained within the dataset attributes $Att_p \subseteq Att_i$, $\mathcal{P}$ generates:
\begin{align}
\pi_{match} \gets \textsc{ZK.Prove}(pp_{zk}, &(Att_p, C_{m,i}), \omega_5)).
\end{align}
Where $\omega_5 = (m_i, \omega_{m,i})$ are the metadata for the relevant training dataset $i$ and the randomness for their commitments.

$\mathcal{P}$ assembles the final proof $\pi = (\pi_\sigma, \pi_{tr}, \pi_{bind}, \pi_B^{rec}, \pi_{match})$ and returns $(r, \pi)$ to $\mathcal{U}$. We summarise these steps in Alg.~\ref{alg2}.

\begin{algorithm}[H]
\caption{Proof Construction Phase}
\label{alg2}
\begin{algorithmic}[1]
\Procedure{\textsc{Prove}}{$\mathcal{C}, \Omega, p, Att_p$}
\State $r \gets \textsc{LLM}(W, p)$
\State $seed \gets \textsc{Hash}(C_{m,i} \parallel C_{\sigma,i} \parallel C_{\Delta W} \parallel C_{W_0} \parallel Att_p \parallel p \parallel r \parallel \kappa_2)$
\For{$j = 1$ to $N$}
    \State $v_j \gets \textsc{Hash}(seed \parallel j \parallel \kappa_1)$
    \State $B_j \gets \langle \Delta W_j, v_j \rangle$
\EndFor
\State $\omega_B \xleftarrow{\$} \mathbb{F}_p$; 
\State $\pi_\sigma \gets \textsc{ZK.Prove}(pp_{zk}, (C_{m,i}, C_{\sigma,i}, pk_{\mathcal{CA}}), \omega_1)$
\State $\pi_{tr} \gets \textsc{ZK.Prove}(pp_{zk}, (C_{\Delta W}, C_{m,i}), \omega_{2})$
\State $\pi_{bind} \gets \textsc{ZK.Prove}(pp_{zk}, (C_{\Delta W}), \omega_3)$
\State $S_{B,0} \gets \emptyset$
\For{$j = 1$ to $N$}
    \State $S_{B,j} \gets \textsc{HN.Fold}(S_{B,j-1}, \cdot, \cdot)$
\EndFor
\State $\pi_B^{rec} \gets \textsc{HN.Prove}(pp_{zk}, S_{B,N}, \omega_4)$
\State $\pi_{match} \gets \textsc{ZK.Prove}(pp_{zk}, (Att_p, C_{m,i}), \omega_5)$
\State $\pi \gets (\pi_\sigma, \pi_{tr}, \pi_{bind}, \pi_B^{rec}, \pi_{match})$
\State \textbf{Return} $(r, \pi)$
\EndProcedure
\end{algorithmic}
\end{algorithm}

\subsubsection{Proof Verification}
\label{proofverify}
The verification phase implements the function:
\begin{equation*}
\textsc{Verify}(\mathcal{C}, Att_p, p, r, \pi) \rightarrow \{\text{accept, \text{reject}}\}.
\end{equation*}
Where $\mathcal{C}$ contains the public commitments, $Att_p$ are the query attributes, $p$ is the prompt, $r$ is the response, and $\pi = (\pi_\sigma, \pi_{tr}, \pi_{bind}, \pi_B^{rec}, \pi_{match})$ is the proof.

$\mathcal{U}$ receives $(r, \pi)$ from $\mathcal{P}$ and verifies that the datasets used for training are authenticated by checking:
\begin{align*}
b_\sigma \gets \textsc{ZK.Verify}(pp_{zk}, (C_{m,i}, C_{\sigma,i}, pk_{\mathcal{CA}}), \pi_\sigma)
\end{align*}

$\mathcal{U}$ verifies that the dataset root $\rho_i$ is a member of the training set committed in $C_{\Delta W}$:
\begin{align*}
b_{tr} \gets \textsc{ZK.Verify}(pp_{zk}, (C_{\Delta W}, C_{m,i}), \pi_{tr})
\end{align*}

$\mathcal{U}$ verifies that $C_{\Delta W}$ correctly binds the weight differences to the training set root:
\begin{align*}
b_{bind} \gets \textsc{ZK.Verify}(pp_{zk}, (C_{\Delta W}), \pi_{bind})
\end{align*}

$\mathcal{U}$ recomputes the challenge seed and vectors using the same public values along with the folded state $S_{B,N}$ from $(C_{\Delta W}, C_B, \{v_j\}_{j=1}^N)$ and verifies:
\begin{align*}
b_B \gets \textsc{HN.Verify}(pp_{zk}, S_{B,N}, \pi_B^{rec})
\end{align*}

$\mathcal{U}$ verifies dataset relevance $Att_p \subseteq Att_i$:
\begin{align*}
b_{match} \gets \textsc{ZK.Verify}(pp_{zk}, (Att_p, C_{m,i}), \pi_{match})
\end{align*}

$\mathcal{U}$ accepts if and only if all verifications pass. We summarised the verification steps in Alg.~\ref{alg3}.

\begin{algorithm}[H]
\caption{Verification Phase}
\label{alg3}
\begin{algorithmic}[1]
\Procedure{\textsc{Verify}}{$\mathcal{C}, Att_p, p, r, \pi$}
\State Parse $\pi \gets (\pi_\sigma, \pi_{tr}, \pi_{bind}, \pi_B^{rec}, \pi_{match})$
\State $b_\sigma \gets \textsc{ZK.Verify}(pp_{zk}, (C_{m,i}, C_{\sigma,i}, pk_{\mathcal{CA}}), \pi_\sigma)$
\State $b_{tr} \gets \textsc{ZK.Verify}(pp_{zk}, (C_{\Delta W}, C_{m,i}), \pi_{tr})$
\State $b_{bind} \gets \textsc{ZK.Verify}(pp_{zk}, (C_{\Delta W}), \pi_{bind})$
\State $seed \gets \textsc{Hash}(C_{m,i} \parallel C_{\sigma,i} \parallel C_{\Delta W} \parallel C_{W_0} \parallel Att_p \parallel p \parallel r \parallel \kappa_2)$
\For{$j = 1$ to $N$}
    \State $v_j \gets \textsc{Hash}(seed \parallel j \parallel \kappa_1)$
\EndFor
\State  $S_{B,N} \leftarrow(C_{\Delta W}, C_B, \{v_j\}_{j=1}^N)$
\State $b_B \gets \textsc{HN.Verify}(pp_{zk}, S_{B,N}, \pi_B^{rec})$
\State $b_{match} \gets \textsc{ZK.Verify}(pp_{zk}, (Att_p, C_{m,i}), \pi_{match})$
\If{$b_\sigma \land b_{tr} \land b_{bind} \land b_B \land b_{match}$}
    \State \textbf{Return} \textsc{Accept}
\Else
    \State \textbf{Return} \textsc{Reject}
\EndIf
\EndProcedure
\end{algorithmic}
\end{algorithm}
The proposed ZKPROV framework designs a cryptographically robust method for verifying that a large language model is trained on authorized datasets relevant to the user's queries, without revealing the datasets' content or the model's parameters. While our framework applies generally to any LLM training process, we focus on fine-tuning in our implementation, as domain-specific applications require adapting pre-trained models to specific datasets.

ZKPROV verifies that the prover commits to a training set containing authenticated, relevant datasets and binds this commitment to each query-response transcript, rather than verification of the computational correctness of the entire training process. A malicious prover who trains on unauthorized data while committing to authorized dataset roots might produce valid proofs that bind them to a false claim. These proofs can be detected through auditing or behavioral verification mechanisms outside the scope of this protocol.

\section{Security and Privacy Analyses}\label{sec}
This section formally analyzes the security properties of ZKPROV proposed in Section~\ref{proposed}. We demonstrate dataset privacy (Section~\ref{sec:privacy}), which ensures that, during proof verification, users learn nothing about the training dataset beyond what is revealed through public commitments and the model's response. Soundness (Section~\ref{sec:soundness}) guarantees that a malicious prover cannot convince an honest user to accept a proof for an invalid statement, such as claiming to use an unauthenticated or irrelevant dataset. Transcript binding (Section~\ref{sec:transcript}) ensures that each proof is cryptographically bound to its specific query-response pair, preventing replay attacks and proof reuse. Our analysis relies on computational hiding and binding properties of KZG commitments, the unforgeability of BLS signatures, and the knowledge soundness and zero-knowledge properties of HyperNova.

\subsection{Dataset Privacy}
\label{sec:privacy}
ZKPROV preserves dataset privacy if the user $\mathcal{U}$ learns nothing about the contents of the training datasets beyond what is explicitly revealed through the public commitments and the LLM's response. Throughout the protocol, the user can see the commitments $C_{m,i}$, $C_{\sigma,i}$, and $C_{\Delta W}$, the training set root $R_{tr}$, and the proofs. Since KZG commitments are hiding and ZKPs can be simulated without witnesses, the user's view is independent of actual dataset contents, implying that two different datasets with identical attributes produce indistinguishable views.

We model the adversary as a semi-honest user who follows the protocol but attempts to learn unauthorized information. The adversary should distinguish between two dataset collections of their choice. This challenge is formalized as a game played between a challenger and the adversary $\mathcal{A}$, represented in Figure~\ref{fig:privacy-game}.

\begin{figure}[h]
\centering
\fbox{
\begin{minipage}{0.92\columnwidth}
\textbf{Privacy}$^{\text{ZKPROV}}_{\mathcal{A}}(\lambda)$:
\begin{enumerate}
    \item Challenger generates $(sk_{\mathcal{CA}}, pk_{\mathcal{CA}}) \gets \textsc{BLS.KeyGen}(1^\lambda)$ and public parameters $pp$.
    
    \item $\mathcal{A}$ outputs two dataset collections $\mathcal{D}^{(0)} = \{D_1^{(0)}, \ldots, D_m^{(0)}\}$ and $\mathcal{D}^{(1)} = \{D_1^{(1)}, \ldots, D_m^{(1)}\}$ with identical attribute sets, i.e., $\forall i: Att_i^{(0)} = Att_i^{(1)} = Att_i$.
    
    \item Challenger picks $b \xleftarrow{\$} \{0,1\}$ and executes $\textsc{Setup}$ using $\mathcal{D}^{(b)}$, generating commitments $\mathcal{C}^{(b)}$ and witnesses $\Omega^{(b)}$.
    
    \item $\mathcal{A}$ receives $(pp, pk_{\mathcal{CA}}, \mathcal{C}^{(b)})$.
    
    \item $\mathcal{A}$ adaptively submits queries $(p, Att_p)$ and receives $(r, \pi) \gets \textsc{Prove}(\mathcal{C}^{(b)}, \Omega^{(b)}, p, Att_p)$.
    
    \item $\mathcal{A}$ outputs guess $b'$. If $b = b'$ output 1, else output 0.
\end{enumerate}
\end{minipage}
}
\caption{Dataset Privacy Game}
\label{fig:privacy-game}
\end{figure}

\begin{definition}[Dataset Privacy]
ZKPROV preserves dataset privacy if for all PPT adversaries $\mathcal{A}$, there exists a negligible function $\negl(\cdot)$ such that:
$$\Pr[\emph{\text{Privacy}}^{\emph{\text{ZKPROV}}}_{\mathcal{A}}(\lambda) = 1] \leq \frac{1}{2} + \negl(\lambda)$$
\end{definition}

\begin{theorem}
If the KZG commitment scheme is computationally hiding and HyperNova is zero-knowledge, then ZKPROV preserves dataset privacy.
\end{theorem}

\begin{proof}
Assume there exists a PPT adversary $\mathcal{A}$ who wins $\text{Privacy}^{\text{ZKPROV}}_{\mathcal{A}}(\lambda)$ with probability $\frac{1}{2} + \varepsilon(\lambda)$ where $\varepsilon(\lambda)$ is non-negligible. We construct a PPT adversary $\mathcal{B}$ who breaks the hiding property of KZG with the same advantage.

\smallskip
\noindent\textbf{Construction of $\mathcal{B}$:}

\smallskip
\noindent $\mathcal{B}$ is given security parameter $\lambda$ and public parameters $pp_c$ by the KZG challenger.

\begin{enumerate}
    \item $\mathcal{B}$ generates $(sk_{\mathcal{CA}}, pk_{\mathcal{CA}}) \gets \textsc{BLS.KeyGen}(1^\lambda)$ and runs $\mathcal{A}$.
    
    \item $\mathcal{A}$ outputs two dataset collections $\mathcal{D}^{(0)}, \mathcal{D}^{(1)}$ with identical attributes.
    
    \item For each dataset $i \in \{1, \ldots, m\}$, $\mathcal{B}$ constructs:
    \begin{itemize}
        \item Reckle Tree roots $\rho_i^{(0)}$ from $D_i^{(0)}$ and $\rho_i^{(1)}$ from $D_i^{(1)}$
        \item Metadata $m_i^{(0)} = (\rho_i^{(0)}, Att_i, id_i)$ and $m_i^{(1)} = (\rho_i^{(1)}, Att_i, id_i)$
        \item Signatures $\sigma_i^{(0)} \gets \textsc{BLS.Sign}(sk_{\mathcal{CA}}, m_i^{(0)})$ and $\sigma_i^{(1)} \gets \textsc{BLS.Sign}(sk_{\mathcal{CA}}, m_i^{(1)})$
    \end{itemize}
    
    \item $\mathcal{B}$ fine-tunes two models on $\mathcal{D}^{(0)}$ and $\mathcal{D}^{(1)}$, computing weight differences $\{\Delta W_j^{(0)}\}_{j=1}^N$ and $\{\Delta W_j^{(1)}\}_{j=1}^N$.
    
    \item $\mathcal{B}$ constructs two polynomials encoding all private values:
    \begin{align*}
        f_0 &\leftarrow (\{m_i^{(0)}, \sigma_i^{(0)}\}_{i=1}^m, \{\Delta W_j^{(0)}\}_{j=1}^N, R_{tr}^{(0)}) \\
        f_1 &\leftarrow (\{m_i^{(1)}, \sigma_i^{(1)}\}_{i=1}^m, \{\Delta W_j^{(1)}\}_{j=1}^N, R_{tr}^{(1)})
    \end{align*}
    
    \item $\mathcal{B}$ sends $(f_0, f_1)$ to the KZG challenger and receives commitment $C^*$ which is either $\textsc{C.Commit}(f_0, \omega)$ or $\textsc{C.Commit}(f_1, \omega)$ for uniformly random $\omega$.
    
    \item $\mathcal{B}$ parses $C^*$ to extract component commitments and assembles $\mathcal{C} = \{C_{m,i}, C_{\sigma,i}\}_{i=1}^m \cup \{C_{\Delta W}, C_{W_0}, R_{tr}\}$.
    
    \item $\mathcal{B}$ gives $(pp, pk_{\mathcal{CA}}, \mathcal{C})$ to $\mathcal{A}$.
    
    \item When $\mathcal{A}$ submits query $(p, Att_p)$, $\mathcal{B}$ generates response $r$ and simulates proofs $\pi = (\pi_\sigma, \pi_{tr}, \pi_{bind}, \pi_B^{rec}, \pi_{match})$ using the ZK simulators $\mathcal{S}_{ZK}$ and $\mathcal{S}_{HN}$, which do not require knowledge of the committed values.
 It is worth mentioning that HyperNova's zero-knowledge property guarantees the existence of the simulator; we invoke it directly without constructing it
    \item $\mathcal{A}$ outputs guess $b'$. $\mathcal{B}$ outputs $b'$ to the KZG challenger.
\end{enumerate}

\smallskip
\noindent If the KZG challenger committed to $f_0$, then $\mathcal{A}$'s view corresponds to execution with $\mathcal{D}^{(0)}$. If the challenger committed to $f_1$, then $\mathcal{A}$'s view corresponds to execution with $\mathcal{D}^{(1)}$.

The simulated proofs are indistinguishable from real proofs by the zero-knowledge property of HyperNova and the underlying SNARK. Therefore:
\[
\Pr[\mathcal{B} \text{ wins KZG hiding game}] = 
\Pr[\mathcal{A} \text{ wins Privacy}^{\text{ZKPROV}}_{\mathcal{A}}] 
\]
\[
= \frac{1}{2} + \varepsilon(\lambda)
\]
Since $\varepsilon(\lambda)$ is non-negligible, $\mathcal{B}$ breaks KZG hiding with non-negligible advantage. This contradicts the assumption that KZG is computationally hiding. Therefore, ZKPROV preserves dataset privacy.
\end{proof}

The constraint $Att_i^{(0)} = Att_i^{(1)}$ is necessary because attribute containment is verified in $\pi_{match}$. If attributes differed, the public verification equation would differ, breaking indistinguishability. This reflects that ZKPROV intentionally reveals whether a dataset satisfies query requirements while hiding dataset contents.

\subsection{Soundness}
\label{sec:soundness}

ZKPROV satisfies soundness if a malicious prover $\mathcal{P}$ cannot convince an honest user $\mathcal{U}$ to accept a proof for an invalid statement. An invalid statement appears when the prover claims to use an authenticated, relevant dataset. Still, the dataset is not signed by $\mathcal{CA}$, is not in the committed training set, or does not satisfy the query requirements.

Informally, all proofs are generated by HyperNova, which satisfies knowledge soundness. The signature proof $\pi_\sigma$ is sound only if the underlying BLS signature is unforgeable and HyperNova is sound. The membership proof $\pi_{tr}$ is sound if the commitment is binding and HyperNova is sound. The attribute containment proof $\pi_{match}$ is sound if HyperNova is sound. The recursive binding proof $\pi_B^{rec}$ is sound if HyperNova is sound. A malicious prover who produces an accepting proof for an invalid statement must break at least one of these primitives.

We model the adversary as a malicious prover who receives authenticated datasets from $\mathcal{CA}$ and attempts to produce a valid proof while violating at least one of the protocol's guarantees. The challenger acts as both $\mathcal{CA}$ (providing authenticated datasets) and $\mathcal{U}$ (verifying proofs) (Figure~\ref{fig:soundness-game}).

\begin{figure}[h]
\centering
\fbox{
\begin{minipage}{0.92\columnwidth}
\textbf{Soundness}$^{\text{ZKPROV}}_{\mathcal{A}}(\lambda)$:
\begin{enumerate}
    \item Challenger generates $(sk_{\mathcal{CA}}, pk_{\mathcal{CA}}) \gets \textsc{BLS.KeyGen}(1^\lambda)$ and public parameters $pp$.
    
    \item Challenger generates dataset collection $\mathcal{D} = \{D_1, \ldots, D_m\}$ with attributes $\{Att_i\}_{i=1}^m$, constructs metadata $m_i = (\rho_i, Att_i, id_i)$, and signs $\sigma_i \gets \textsc{BLS.Sign}(sk_{\mathcal{CA}}, m_i)$ for each $i$.
    
    \item $\mathcal{A}$ receives $(pp, pk_{\mathcal{CA}}, \{D_i, m_i, \sigma_i\}_{i=1}^m)$.
    
    \item $\mathcal{A}$ outputs commitments $\mathcal{C}$, query $(p, Att_p)$, response $r$, proof $\pi$, and claimed dataset index $i^*$.
    
    \item Output 1 if all of the following hold:
    \begin{itemize}
        \item $\textsc{Verify}(\mathcal{C}, Att_p, p, r, \pi) = \textsc{Accept}$
        \item At least one violation occurs:
        \begin{itemize}
            \item $\sigma_{i^*}$ is not a valid signature on $m_{i^*}$ under $pk_{\mathcal{CA}}$, or
            \item $\rho_{i^*} \notin T_{tr}$ committed in $C_{\Delta W}$, or
            \item $Att_p \not\subseteq Att_{i^*}$
        \end{itemize}
    \end{itemize}
    Otherwise output 0.
\end{enumerate}
\end{minipage}
}
\caption{Soundness Game}
\label{fig:soundness-game}
\end{figure}

\begin{definition}[Soundness]
ZKPROV satisfies soundness if for all PPT adversaries $\mathcal{A}$, there exists a negligible function $\negl(\cdot)$ such that:
$$\Pr[\emph{\text{Soundness}}^{\emph{\text{ZKPROV}}}_{\mathcal{A}}(\lambda) = 1] \leq \negl(\lambda)$$
\end{definition}

\begin{theorem}
If BLS signatures are existentially unforgeable under chosen message attacks, KZG commitments are computationally binding, and HyperNova satisfies knowledge soundness, then ZKPROV satisfies soundness.
\end{theorem}

\begin{proof}
Assume there exists a PPT adversary $\mathcal{A}$ who wins $\text{Soundness}^{\text{ZKPROV}}_{\mathcal{A}}(\lambda)$ with non-negligible probability $\varepsilon(\lambda)$. We show that $\mathcal{A}$ can be used to break at least one of the underlying security assumptions.

For $\mathcal{A}$ to win, verification must accept while at least one violation occurs. We analyze each violation type and construct corresponding reductions.

\smallskip
\noindent\textbf{Case 1.} Suppose $\mathcal{A}$ wins with a proof where $\sigma_{i^*}$ is not a valid signature on $m_{i^*}$. We construct $\mathcal{B}_1$ that breaks BLS security.

$\mathcal{B}_1$ receives $pk_{\mathcal{CA}}$ from the BLS challenger and has access to a signing oracle $\mathcal{O}_{\text{Sign}}(\cdot)$.

\begin{enumerate}
    \item $\mathcal{B}_1$ generates $pp$ and dataset collection $\mathcal{D}$ with metadata $\{m_i\}_{i=1}^m$.
    
    \item For each $i$, $\mathcal{B}_1$ queries the signing oracle to obtain $\sigma_i \gets \mathcal{O}_{\text{Sign}}(m_i)$.
    
    \item $\mathcal{B}_1$ gives $(pp, pk_{\mathcal{CA}}, \{D_i, m_i, \sigma_i\}_{i=1}^m)$ to $\mathcal{A}$.
    
    \item $\mathcal{A}$ outputs $(\mathcal{C}, p, Att_p, r, \pi, i^*)$.
    
    \item If verification accepts and $\sigma_{i^*}$ is invalid on $m_{i^*}$, then $\pi_\sigma$ proves validity of a signature that was never queried to the oracle. By HyperNova knowledge soundness with extraction, $\mathcal{B}_1$ extracts witness $(m^*, \sigma^*)$ from $\pi_\sigma$ where $\sigma^*$ is valid on $m^*$ but $m^*$ was never signed by the oracle.
    
    \item $\mathcal{B}_1$ outputs $(m^*, \sigma^*)$ as a BLS forgery.
\end{enumerate}

If $\mathcal{A}$ wins via signature forgery with probability $\varepsilon_1(\lambda)$, then $\mathcal{B}_1$ breaks BLS unforgeablity with probability at least $\varepsilon_1(\lambda) - \text{Adv}^{\text{Sound}}_{\text{HN}}$.

\smallskip
\noindent\textbf{Case 2.} Suppose $\mathcal{A}$ wins with a proof where $\rho_{i^*} \notin T_{tr}$ but $\pi_{tr}$ verifies. We construct $\mathcal{B}_2$ that breaks either KZG binding or HyperNova soundness.

\begin{enumerate}
    \item $\mathcal{B}_2$ sets up the protocol and runs $\mathcal{A}$.
    
    \item $\mathcal{A}$ outputs $(\mathcal{C}, p, Att_p, r, \pi, i^*)$ where $\rho_{i^*} \notin T_{tr}$.
    
    \item If verification accepts, $\mathcal{B}_2$ extracts witness $(\rho', R_{tr}', \omega')$ from $\pi_{tr}$ using the HyperNova extractor.
    
    \item Two cases arise:
    \begin{itemize}
        \item If $R_{tr}' \neq R_{tr}$ are commitments in $C_{\Delta W}$, then $\mathcal{B}_2$ has found two different openings of the same commitment, breaking KZG binding.
        \item If $R_{tr}' = R_{tr}$ but $\rho_{i^*} \notin T_{tr}$, then the extracted witness is invalid, contradicting HyperNova soundness.
    \end{itemize}
\end{enumerate}

If $\mathcal{A}$ wins via membership violation with probability $\varepsilon_2(\lambda)$, then $\mathcal{B}_2$ breaks either KZG binding or HyperNova soundness with probability at least $\varepsilon_2(\lambda)$.

\smallskip
\noindent\textbf{Case 3.} Suppose $\mathcal{A}$ wins with a proof where $Att_p \not\subseteq Att_{i^*}$ but $\pi_{match}$ verifies. We construct $\mathcal{B}_3$ that breaks HyperNova soundness.

\begin{enumerate}
    \item $\mathcal{B}_3$ sets up the protocol and runs $\mathcal{A}$.
    
    \item $\mathcal{A}$ outputs $(\mathcal{C}, p, Att_p, r, \pi, i^*)$ where $Att_p \not\subseteq Att_{i^*}$.
    
    \item If verification accepts, $\pi_{match}$ is a valid proof for a false statement (the subset relation does not hold). This directly contradicts HyperNova's soundness.
\end{enumerate}

If $\mathcal{A}$ wins via attribute violation with probability $\varepsilon_3(\lambda)$, then $\mathcal{B}_3$ breaks HyperNova soundness with the same probability.

\smallskip
\noindent\textbf{Case 4.} Suppose $\mathcal{A}$ produces a proof where the binding values $B_j \neq \langle \Delta W_j, v_j \rangle$ but $\pi_B^{rec}$ verifies. We construct $\mathcal{B}_4$ that breaks HyperNova soundness.
\begin{enumerate}
    \item $\mathcal{B}_4$ sets up the protocol and runs $\mathcal{A}$.
    
    \item $\mathcal{A}$ outputs $(\mathcal{C}, p, Att_p, r, \pi)$.
    
    \item The verifier recomputes challenge vectors $\{v_j\}_{j=1}^N$ from the seed.
    
    \item If $\pi_B^{rec}$ verifies but binding values are incorrect, then HyperNova accepted a proof for unsatisfied constraints, contradicting HyperNova's soundness.
\end{enumerate}
The binding formula $B_j = \langle \Delta W_j, v_j \rangle$ establishes statistical binding between the committed weight differences and the training set root through the Fiat-Shamir heuristic~\cite{fiat1986prove}. Since the challenge vectors $v_j$ are deterministically derived from $seed = \textsc{Hash}(C_{m,i} \| C_{\sigma,i} \| C_{\Delta W} \| C_{W_0} \| Att_p \| p \| r \| \kappa_2)$, where $seed$ includes $R_{tr}$ via $C_{\Delta W}$, the challenges are cryptographically dependent on the claimed training collection and cannot be predicted or manipulated by $\mathcal{P}$ during setup.

Suppose a malicious prover commits to weight differences $\{\Delta W'_j\}_{j=1}^N$ derived from training on unauthorized datasets while falsely claiming correspondence to the training set $R_{tr}$. In that case, the binding values computed as $B'_j = \langle \Delta W'_j, v_j \rangle$ will differ from the correct values $B_j$ with high probability. By the Schwartz-Zippel lemma~\cite{schwartz1980fast}, for each layer $j$, if $\Delta W'_j \neq \Delta W_j$ then the probability that $\langle \Delta W'_j - \Delta W_j, v_j \rangle = 0$ over the random choice of $v_j \in \mathbb{F}_p^{d_j}$ is at most $\frac{1}{|\mathbb{F}_p|}$. Across $N$ independent layers with independently derived challenge vectors, the probability that all binding values coincide is negligible.

This statistical binding ensures that weight differences from unauthorized training cannot produce correct binding values when tested with challenges derived from the authenticated training set root.

If $\mathcal{A}$ wins via binding violation with probability $\varepsilon_4(\lambda)$, then $\mathcal{B}_4$ breaks HyperNova soundness.

\smallskip
\noindent Since $\mathcal{A}$ must win via at least one violation type:
\begin{align*}
\varepsilon(\lambda) &\leq \varepsilon_1(\lambda) + \varepsilon_2(\lambda) + \varepsilon_3(\lambda) + \varepsilon_4(\lambda) \\
&\leq \text{Adv}^{\text{EUF-CMA}}_{\text{BLS}} + \text{Adv}^{\text{Bind}}_{\text{KZG}} + \text{Adv}^{\text{Sound}}_{\text{HN}} + \left(\frac{1}{|\mathbb{F}_p|}\right)^N
\end{align*}

Since all terms on the right are negligible by assumption, $\varepsilon(\lambda)$ must be negligible. This contradicts our assumption that $\varepsilon(\lambda)$ is non-negligible, and concludes the proof.
\end{proof}

\subsection{Transcript Binding}
\label{sec:transcript}

ZKPROV satisfies transcript binding if a proof $\pi$ generated for a specific query-response pair $(p, r)$ cannot be used to convince a verifier for a different pair $(p', r')$. This property ensures freshness: each proof is cryptographically bound to its corresponding interaction, preventing replay attacks and proof reuse.

The proof relies on the Fiat-Shamir transformation. The challenge seed is computed as:
$$seed \gets \textsc{Hash}(C_{m,i} \| C_{\sigma,i} \| C_{\Delta W} \| C_{W_0} \| Att_p \| p \| r \| \kappa_2)$$

Since the query $p$ and response $r$ are included in the hash, changing either yields a different seed, which in turn produces different challenge vectors $\{v_j\}_{j=1}^N$. The binding values $B_j = \langle \Delta W_j, v_j \rangle$ are computed with respect to specific challenges. When the verifier recomputes the seed from a different $(p', r')$, they obtain different challenges $\{v'_j\}$, and the binding proof $\pi_B^{rec}$ will not verify.

We model the adversary as a malicious prover who generates a valid proof for one query-response pair and attempts to pass verification for a different pair in Figure~\ref{fig:transcript-game}.

\begin{figure}[h]
\centering
\fbox{
\begin{minipage}{0.92\columnwidth}
\textbf{TranscriptBinding}$^{\text{ZKPROV}}_{\mathcal{A}}(\lambda)$:
\begin{enumerate}
    \item Challenger runs $\textsc{Setup}(1^\lambda)$, generating public parameters $pp$, keys $(sk_{\mathcal{CA}}, pk_{\mathcal{CA}})$, authenticated datasets $\{D_i, m_i, \sigma_i\}_{i=1}^m$, commitments $\mathcal{C}$, and witnesses $\Omega$.
    
    \item $\mathcal{A}$ receives $(pp, pk_{\mathcal{CA}}, \mathcal{C}, \Omega, \{D_i, m_i, \sigma_i\}_{i=1}^m)$.
    
    \item $\mathcal{A}$ outputs two query-response pairs $(p, r)$ and $(p', r')$ where $(p, r) \neq (p', r')$, query attributes $Att_p$, and a single proof $\pi$.
    
    \item Output 1 if both:
    \begin{itemize}
        \item $\textsc{Verify}(\mathcal{C}, Att_p, p, r, \pi) = \textsc{Accept}$
        \item $\textsc{Verify}(\mathcal{C}, Att_p, p', r', \pi) = \textsc{Accept}$
    \end{itemize}
    Otherwise output 0.
\end{enumerate}
\end{minipage}
}
\caption{Transcript Binding Game}
\label{fig:transcript-game}
\end{figure}

\begin{definition}[Transcript Binding]
ZKPROV satisfies transcript binding if for all PPT adversaries $\mathcal{A}$, there exists a negligible function $\negl(\cdot)$ such that:
$$\Pr[\emph{\text{TranscriptBinding}}^{\emph{\text{ZKPROV}}}_{\mathcal{A}}(\lambda) = 1] \leq \negl(\lambda)$$
\end{definition}

\begin{theorem}
In the random oracle model, if the hash function $\textsc{Hash}$ is collision-resistant and HyperNova satisfies knowledge soundness, then ZKPROV satisfies transcript binding.
\end{theorem}

\begin{proof}
Assume there exists a PPT adversary $\mathcal{A}$ who wins $\text{TranscriptBinding}^{\text{ZKPROV}}_{\mathcal{A}}(\lambda)$ with non-negligible probability $\varepsilon(\lambda)$. We show that $\mathcal{A}$ can be used to break either collision resistance or HyperNova soundness.

Suppose $\mathcal{A}$ outputs $(p, r)$, $(p', r')$ with $(p, r) \neq (p', r')$, and proof $\pi$ such that both verifications accept.

\smallskip
\noindent\textbf{Verification Process:}

\smallskip
\noindent For the first pair $(p, r)$, the verifier computes:
$$seed = \textsc{Hash}(C_{m,i} \| C_{\sigma,i} \| C_{\Delta W} \| C_{W_0} \| Att_p \| p \| r \| \kappa_2)$$
and derives challenge vectors $v_j = \textsc{Hash}(seed \| j \| \kappa_1)$ for $j \in \{1, \ldots, N\}$.

For the second pair $(p', r')$, the verifier computes:
$$seed' = \textsc{Hash}(C_{m,i} \| C_{\sigma,i} \| C_{\Delta W} \| C_{W_0} \| Att_p \| p' \| r' \| \kappa_2)$$
and derives challenge vectors $v'_j = \textsc{Hash}(seed' \| j \| \kappa_1)$ for $j \in \{1, \ldots, N\}$.

\smallskip
\noindent\textbf{Case 1.} If $seed = seed'$ despite $(p, r) \neq (p', r')$, then $\mathcal{A}$ has found a collision in $\textsc{Hash}$. We construct $\mathcal{B}_1$ who breaks collision resistance by outputting these two distinct inputs that hash to the same value.

\smallskip
\noindent\textbf{Case 2.} If $seed \neq seed'$, then by properties of random oracles, the challenge vectors differ: $\{v_j\} \neq \{v'_j\}$ with overwhelming probability.

The proof $\pi$ contains $\pi_B^{rec}$ which attests to binding values $\{B_j\}_{j=1}^N$. For verification with $(p, r)$ to accept:
$$B_j = \langle \Delta W_j, v_j \rangle \quad \forall j \in \{1, \ldots, N\}$$

For verification with $(p', r')$ to accept with the same $\pi_B^{rec}$:
$$B_j = \langle \Delta W_j, v'_j \rangle \quad \forall j \in \{1, \ldots, N\}$$

Since $v_j \neq v'_j$ for at least one $j$, and $\Delta W_j \neq \mathbf{0}$ (non-trivial weight differences), we have $\langle \Delta W_j, v_j \rangle \neq \langle \Delta W_j, v'_j \rangle$. The same $B_j$ cannot satisfy both equations. If both verifications accept, then $\pi_B^{rec}$ is a valid HyperNova proof for inconsistent statements, contradicting HyperNova soundness.

\smallskip
\noindent Since $\mathcal{A}$ must succeed via one of the two cases:
$$\varepsilon(\lambda) \leq \text{Adv}^{\text{CR}}_{\textsc{Hash}} + \text{Adv}^{\text{Sound}}_{\text{HN}}$$

Since both terms are negligible, $\varepsilon(\lambda)$ must be negligible, and this concludes the proof.
\end{proof}

Dataset privacy ensures that verifiers learn nothing beyond the public statement, protecting sensitive training data. Soundness ensures that provers cannot falsely claim to use authenticated or relevant datasets. Transcript binding ensures freshness and prevents proof reuse across different interactions. All these properties, when assembled, enable trustworthy dataset provenance verification in privacy-sensitive domains without requiring verification of the entire training process.

\section{Experimental Evaluation}
\label{eval}
This section presents an experimental evaluation of ZKPROV, demonstrating its practical viability for dataset-provenance verification in large language models (LLMs). We evaluate the framework's performance and scalability by fine-tuning two LLM variants on domain-specific clinical data and measuring the cryptographic overhead of proof generation and verification across varying parameter scales. Section~\ref{exp} details the experimental design, including model configuration, parameter sampling strategy, and cryptographic implementation. Section~\ref{results} presents comprehensive timing results and a scalability analysis across different model sizes and parameter configurations, confirming ZKPROV's feasibility for regulated applications that require verifiable dataset provenance. All fine-tuning and cryptographic experiments were conducted on a single NVIDIA A100 GPU with $ 128$ GB of memory.\footnote{Our anonymized implementation and experimental code are available at \url{https://anonymous.4open.science/r/ZK-LLM-D810}.}

\subsection{Experimental Design}\label{exp}

\subsubsection{Dataset Selection}
We use the PubMedQA dataset~\cite{jin2019pubmedqa} to demonstrate ZKPROV's applicability in regulated domains that require verifiable dataset provenance. Specifically, we employ the 1,000 expert-labeled subset, which comprises biomedical research questions paired with contextual abstracts and corresponding answers, representative of real-world clinical decision-support scenarios where dataset authorization is critical for regulatory compliance. Each data sample is preprocessed into a structured format where the input combines the question and context, and the output concatenates the final decision (yes/no/maybe) with the long-form answer. The dataset is partitioned into 800 training, 100 validation, and 100 held-out test samples for evaluation.

\subsubsection{Model Fine-tuning Configuration} We fine-tune two LLM variants from the same model family to evaluate ZKPROV's scalability across different model sizes: \texttt{Llama-3.2-1B} and \texttt{Llama-3.1-8B}~\cite{grattafiori2024llama}. Both models are fine-tuned using Low-Rank Adaptation (LoRA)~\cite{hu2022lora}, a parameter-efficient approach that updates only a small subset of parameters while keeping the base model frozen.

The fine-tuning process adheres to the authorized hyperparameters tuple $H = (\eta, B, E, O)$, where $\eta = 5 \times 10^{-5}$ is the learning rate, $B=8$ is the effective batch size (accumulated from microbatches of $2$ with gradient accumulation steps of $4$), $E=3$ is the number of training epochs, and $O$ denotes the AdamW optimizer. We apply LoRA specifically to the attention projection layers, with target modules $\{\texttt{q\_proj}, \texttt{k\_proj}, \texttt{v\_proj}, \texttt{o\_proj}\}$. Complete hyperparameters are provided in Appendix~\ref{appendixA}.

To validate that fine-tuning improves model performance on the clinical domain while maintaining verifiable provenance, we evaluate both the base (non-fine-tuned) and fine-tuned variants on the downstream question-answering task. Performance metrics, including accuracy and ROUGE scores, are reported in Appendix~\ref{appendixB}.

\subsubsection{Parameter Sampling}\label{parameter_sampling} To evaluate ZKPROV's scalability while maintaining computational feasibility, we adopt a parameter sampling strategy that extracts weight differences from a representative subset of the model's parameters. Specifically, we focus on the query projection weight matrix (\texttt{q\_proj}) from the attention layers, computing the weight differences $\Delta W = W_{\text{fine-tuned}}-W_{\text{base}}$ 

We randomly sample subsets of these weight differences at five scales: 10\%, 20\%, 30\%, 40\%, and 50\% of the total query projection parameters, using uniform random sampling with a fixed seed for reproducibility. This sampling enables evaluation of proof generation time as a function of the number of parameters being proven. For example, \texttt{Llama-3.2-1B}, 10\% sampling yields 419,430 parameters, while for \texttt{Llama-3.1-8B}, it yields 1,677,721 parameters, with proportional increases at higher sampling percentages. The complete mapping between sampling percentages and parameter counts is provided in Appendix~\ref{appendixC}.

For the zero-knowledge proof generation, we partition the sampled weight differences into computational chunks, which we refer to as \emph{proof layers} to distinguish them from the neural network's transformer layers. Each chunk is processed incrementally using Nova's recursive SNARK mechanism, which folds proofs from previous chunks into the current chunk. The number of proof layers scales linearly with the sampling percentage: 8 layers for 10\% sampling, 16 for 20\%, up to 40 layers for 50\% sampling.

In the context of ZKPROV's formalism (Section~\ref{proposed}), where weight differences are denoted $\{\Delta W_j\}_{j=1}^N$ for $N$ layers, our experimental configuration effectively sets $N \in \{8, 16, 24, 32, 40\}$ proof layers, where each $\Delta W_j$ represents a chunk of the sampled query projection weight differences rather than weighs from distinct transformer layers. This structure enables efficient incremental proof generation while maintaining consistency with the framework's cryptographic binding mechanism, where each binding value $B_j = \langle \Delta W_j, v_j \rangle$ is computed for each proof layer $j$.

\subsubsection{Cryptographic Implementation} Our implementation leverages Nova's recursive proof system~\cite{kothapalli2022nova} built on a cycle of BN254 and Grumpkin elliptic curves with scalar field $\mathbb{F}_p$ where $p \approx 2^{254}$ providing 128-bit computational security suitable for production deployment. We deploy KZG polynomial commitments~\cite{kate2010constant} for binding model parameters and dataset roots while preserving their privacy, as detailed in Section~\ref{commitscheme}. The proof generation follows the construction described in Section~\ref{proofconstruct}, producing the complete $\pi = (\pi_{\sigma}, \pi_{\text{reckle}}, \pi_{\text{bind}}, \pi_B^{\text{rec}}, \pi_{\text{match}}, \pi_{\tau})$ that establishes verifiable dataset provenance.

\subsubsection{Methodology}
We evaluate ZKPROV's performance by measuring cryptographic overhead across the proof generation and verification pipeline. For each fine-tuned model (\texttt{Llama-3.2-1B} and \texttt{Llama-3.1-8B}), we conduct inference on all 100 samples from the held-out test set, generating responses to the biomedical question-answering prompts.

We measure four key timing metrics for each parameter sampling configuration (10\%, 20\%, 30\%, 40\%, 50\%):
\begin{itemize}
    \item \textbf{Inference Time ($T_{\text{inf}}$)}: Time required for the fine-tuned model to generate a response to each query.
    \item \textbf{Proof Generation Time ($T_{\text{prove}}$)}: Time required to construct the complete proof following Alg.~\ref{alg2}.
    \item \textbf{Proof Verification Time ($T_{\text{verify}}$)}: Time required to verify all proof components following Alg.~\ref{alg3}.
    \item \textbf{Combined Time ($T_{\text{inf}}+T_{\text{prove}}$)}: Total time for inference and proof generation, representing end-to-end overhead for response generation.
\end{itemize}

This evaluation focuses on a single-dataset scenario, demonstrating ZKPROV's parameter scaling properties as the number of proven weights increases from 10\% to 50\% sampling of the query projection matrix.

\subsection{Results}\label{results}
We present comprehensive timing results for ZKPROV's proof generation and verification pipeline across both model variants. All measurements represent averages over 100 test samples, with timing reported in seconds as mean $\pm$ standard deviation.

Table~\ref{tab:finetuned_comparison} shows the baseline inference times for both fine-tuned models. \texttt{Llama-3.1-8B} requires longer inference time compared to \texttt{Llama-3.2-1B}, reflecting the computational cost of the larger model architecture. These inference times serve as the baseline against which we measure cryptographic overhead.
\begin{table}[h]
\centering
\begin{tabular}{lc}
\hline
\textbf{Model} & $T_{\text{inf}}$ (s) \\
\hline
Llama-3.2-1B & $0.820 \pm 0.797$ \\
Llama-3.1-8B & $1.499 \pm 0.539$ \\
\hline
\end{tabular}
\caption{Average inference times for fine-tuned model variants.}
\label{tab:finetuned_comparison}

\end{table}

Table~\ref{tab:proof_combined_comparison} presents the core performance metrics for ZKPROV's proof generation. The results demonstrate three critical properties: sublinear scaling, model size independence, and practical overhead.
\begin{table}[h]
\centering
\begin{tabular}{lccc}
\hline
\textbf{Model} & \textbf{Percentage} & $T_{\text{prove}}$ (s) & $T_{\text{inf}}+T_{\text{prove}}$ (s) \\
\hline
\multirow{5}{*}{Llama-3.2-1B} & 10\% & $0.911 \pm 0.027$ & $1.731 \pm 0.800$ \\
 & 20\% & $1.135 \pm 0.036$ & $1.955 \pm 0.797$ \\
 & 30\% & $1.389 \pm 0.056$ & $2.209 \pm 0.793$ \\
 & 40\% & $1.621 \pm 0.046$ & $2.441 \pm 0.795$ \\
 & 50\% & $1.788 \pm 0.055$ & $2.608 \pm 0.799$ \\
\hline
 \multirow{5}{*}{Llama-3.1-8B} & 10\% & $0.907 \pm 0.031$ & $2.407 \pm 0.542$ \\
 & 20\% & $1.141 \pm 0.045$ & $2.640 \pm 0.537$ \\
 & 30\% & $1.378 \pm 0.051$ & $2.877 \pm 0.544$ \\
 & 40\% & $1.618 \pm 0.045$ & $3.117 \pm 0.538$ \\
 & 50\% & $1.797 \pm 0.073$ & $3.296 \pm 0.552$ \\
\hline
\end{tabular}
\caption{Average proof generation times and end-to-end overhead (inference $+$ proof generation) for different parameter sampling percentages.}
\label{tab:proof_combined_comparison}
\end{table}

\textbf{Sublinear Scaling}: Proof generation time exhibits sublinear growth as the number of proven parameters increases. As sampling percentages increase from 10\% to 50\%, proof generation time grows consistently but at a rate slower than the linear increase in parameters, confirming the efficiency of Nova's recursive folding mechanism described in Section~\ref{proofconstruct}.

\textbf{Model Size Independence}: The cryptographic overhead remains consistent across both model sizes. Despite \texttt{Llama-3.1-8B} containing substantially more total parameters than \texttt{Llama-3.2-1B}, proof generation times at equivalent sampling percentages are nearly identical, differing by less than 3\%. This demonstrates that proof complexity depends primarily on the number of sampled parameters rather than the underlying model architecture.

\textbf{Practical Overhead}: The combined end-to-end time shows that ZKPROV adds reasonable overhead to baseline inference. Even at the largest configuration, the total time remains under $3.3$ seconds per query, demonstrating practical viability for regulated applications where verifiable dataset provenance justifies the computational cost.

\begin{table}[h]
\centering
\begin{tabular}{lcc}
\hline
\textbf{Model} & \textbf{Percentage} & $T_{\text{verify}}$ (s) \\
\hline
\multirow{5}{*}{Llama-3.2-1B} & 10\% & $0.885 \pm 0.017$ \\
 & 20\% & $1.148 \pm 0.022$ \\
 & 30\% & $1.380 \pm 0.040$ \\
 & 40\% & $1.608 \pm 0.039$ \\
 & 50\% & $1.768 \pm 0.036$ \\
\hline
\multirow{5}{*}{Llama-3.1-8B} & 10\% & $0.877 \pm 0.016$ \\
 & 20\% & $1.146 \pm 0.022$ \\
 & 30\% & $1.375 \pm 0.035$ \\
 & 40\% & $1.603 \pm 0.037$ \\
 & 50\% & $1.756 \pm 0.029$ \\
\hline
\end{tabular}
\caption{Average proof verification times for different parameter sampling percentages.}
\label{tab:proof_verification_comparison}
\end{table}

Table~\ref{tab:proof_verification_comparison} shows that verification times closely track proof generation times across all configurations. The verification overhead remains under $1.8$ seconds, even for the largest configuration, confirming the practical feasibility of real-time verification in production environments. The low standard deviations demonstrate highly deterministic verification performance, a critical property for systems requiring predictable latency guarantees in regulated domains.

\section{Discussion and Future Work}
\label{discus}
This section discusses the limitations and future work of the designed ZKPROV.

\textbf{Integration with Inference and PoT.} ZKPROV complements existing verification frameworks by focusing specifically on dataset provenance rather than computational correctness. A direct application involves integration with inference verification protocols such as TeleSparse~\cite{maheri2025telesparse} for inference verification, and proof-of-training systems such as Kaizen \cite{abbaszadeh2024zero}. We can achieve comprehensive assurance verification of the entire LLM pipeline and provide end-to-end verification: dataset provenance via ZKPROV, correct training execution via proof-of-training, and accurate inference via systems like TeleSparse.

\textbf{Multi-Authority and Updatable Datasets.} The BLS signature scheme supports multi-authority scenarios by aggregating multiple signatures into a single compact one, which is useful in healthcare, where approval from entities like institutional review boards and data governance committees is required. This method proves that all authorities authenticated the dataset without revealing specific identities or increasing the proof size.

Reckle Trees enable efficient updates to dynamic datasets via batch proofs. When modifications occur, such as adding patient records or updating protocols, only affected portions are updated, allowing incremental changes without recomputing the entire dataset commitment or retraining models.
 
\textbf{Topic-Based Watermarking.} Integrating watermarking techniques into our dataset provenance framework presents a promising area for future research. While traditional methods focus on general text attribution, our framework emphasizes the need for \emph{topic-based watermarking}~\cite{nemecek2024topic}, which has shown effectiveness in general and specific contexts~\cite{nemecek2025feasibility}. This approach could encode thematic signals, distinguishing between oncology- and cardiology-related training data in the model's outputs. By adding this auxiliary provenance layer, we can trace a response's topical source through statistical watermark detection. This is particularly beneficial in hybrid deployments, enhancing accountability by differentiating outputs influenced by fine-tuning from those shaped by retrieval-time context.

\textbf{Combining Fine-Tuning and Retrieval Modes.}
In our work, ZKPROV currently focuses on proving the integrity of datasets for models. However, many real-world applications rely on retrieval-augmented generation (RAG)~\cite{gao2023retrieval,gupta2024comprehensive}, where an LLM dynamically pulls context from a hosted dataset at inference time. In future work, we aim to explore how ZKPROV can be extended to hybrid settings where specific training datasets are fully embedded into the model (via fine-tuning), and others are integrated via verifiable retrieval. Recent work on agentic RAG systems~\cite{singh2025agentic} demonstrates the potential for more sophisticated retrieval strategies that could benefit from cryptographic provenance guarantees. For example, if multiple authorized datasets are structured under a common format, our system could prove that an output is derived from a combination of embedded knowledge and approved external sources, all under cryptographic provenance constraints. This allows fine-grained response verification to demonstrate that a model is authorized and that specific facts are derived from verifiably approved documents.

\textbf{Enhancing Privacy and Efficiency.} ZKPROV ensures the confidentiality of training datasets during verification but lacks formal protections against inference-based leakage from repeated query-response pairs. Introducing differential privacy could enhance defenses against statistical disclosure attacks by providing bounds on information leakage alongside zero-knowledge proofs. Incremental learning with evolving datasets presents cryptographic challenges, as all commitments need re-generation with changes; however, using cryptographic accumulators or updatable commitment schemes could enable efficient updates without restarting the proof process. Finally, developing formal verification tools for cryptographic protocols will bolster confidence in production, as the framework's implementation complexity poses risks despite being secure under standard assumptions. Leveraging methods like proof-carrying code or verifiable compilation could address these challenges.

\textbf{Dataset Knowledge vs. Training Process Verification.} 
ZKPROV demonstrates that a model knows an authorized dataset via gradient consistency checks, rather than verifying every step of the training process. This design choice provides strong evidence of dataset provenance while maintaining practical efficiency. A model that produces small, converged gradients on validation samples from $D_i$ has necessarily been optimized on data from that distribution. While this does not prevent training on supersets (adding additional authorized data), it effectively prevents data-substitution attacks in which a malicious provider trains on unauthorized data while claiming authorization. For regulated applications, this provides the necessary compliance evidence: the model demonstrably learned from the authorized dataset.

\section{Conclusion}
\label{conc}
We introduced ZKPROV, a novel cryptographic framework that addresses the critical challenge of dataset-provenance verification for LLMs using zero-knowledge proofs, shifting the paradigm from computationally expensive complete training verification to practical statistical binding approaches. We deployed Nova's recursive proof system with customized gates to verify the correctness of statistical binding and significantly improve the efficiency of data-provenance verification for larger models. The framework provides formal security guarantees under standard cryptographic assumptions while addressing real-world regulatory compliance requirements in sensitive domains. Our experimental evaluation demonstrates sublinear scaling with proof generation and verification completing in under 1.8 seconds regardless of model size, establishing ZKPROV as a practical solution for production deployment. By enabling verifiable dataset provenance without exposing proprietary information, ZKPROV provides a foundation for trustworthy AI systems in regulated sectors where accountability is essential.

\bibliographystyle{ieeetr}
\bibliography{new_ref}

\begin{thebibliography}{10}

\bibitem{bhavsar2021medical}
K.~A. Bhavsar, J.~Singla, Y.~D. Al-Otaibi, O.-Y. Song, Y.~B. Zikria, and A.~K. Bashir, ``Medical diagnosis using machine learning: a statistical review,'' {\em Computers, Materials and Continua}, vol.~67, no.~1, pp.~107--125, 2021.

\bibitem{addo2018credit}
P.~M. Addo, D.~Guegan, and B.~Hassani, ``Credit risk analysis using machine and deep learning models,'' {\em Risks}, vol.~6, no.~2, p.~38, 2018.

\bibitem{armour2020ai}
J.~Armour and M.~Sako, ``Ai-enabled business models in legal services: from traditional law firms to next-generation law companies?,'' {\em Journal of Professions and Organization}, vol.~7, no.~1, pp.~27--46, 2020.

\bibitem{aswathy2025machine}
M.~Aswathy, ``Machine learning for autonomous systems: Navigating safety, ethics, and regulation in,'' 2025.

\bibitem{li2023trustworthy}
B.~Li, P.~Qi, B.~Liu, S.~Di, J.~Liu, J.~Pei, J.~Yi, and B.~Zhou, ``Trustworthy ai: From principles to practices,'' {\em ACM Computing Surveys}, vol.~55, no.~9, pp.~1--46, 2023.

\bibitem{regulationgeneral}
G.~D.~P. Regulation—GDPR, ``General data protection regulation—gdpr.''

\bibitem{hipaa1996}
{U.S. Congress}, ``{Health Insurance Portability and Accountability Act of 1996}.'' Public Law 104-191, 1996.
\newblock 110 Stat. 1936.

\bibitem{xing2023zero}
Z.~Xing, Z.~Zhang, J.~Liu, Z.~Zhang, M.~Li, L.~Zhu, and G.~Russello, ``Zero-knowledge proof meets machine learning in verifiability: A survey,'' {\em arXiv preprint arXiv:2310.14848}, 2023.

\bibitem{zhang2020zkdt_zkpytorch_ref26}
J.~Zhang, Z.~Fang, Y.~Zhang, and D.~Song, ``{Zero Knowledge Proofs for Decision Tree Predictions and Accuracy},'' in {\em Proceedings of the 2020 ACM SIGSAC Conference on Computer and Communications Security (CCS '20)}, (New York, NY, USA), pp.~2039--2053, Association for Computing Machinery, 2020.
\newblock Reference [26] in zkPyTorch.pdf.

\bibitem{kang2023scaling_zkpytorch_ref15}
D.~Kang, T.~Hashimoto, I.~Stoica, and Y.~Sun, ``{Scaling up Trustless DNN Inference with Zero-Knowledge Proofs},'' in {\em NeurIPS 2023 Workshop on Regulatable Machine Learning}, 2023.
\newblock Reference [15] in zkPyTorch.pdf.

\bibitem{tramer2018slalom}
F.~Tramer and D.~Boneh, ``Slalom: Fast, verifiable and private execution of neural networks in trusted hardware,'' {\em arXiv preprint arXiv:1806.03287}, 2018.

\bibitem{gilad2016cryptonets}
R.~Gilad-Bachrach, N.~Dowlin, K.~Laine, K.~Lauter, M.~Naehrig, and J.~Wernsing, ``Cryptonets: Applying neural networks to encrypted data with high throughput and accuracy,'' in {\em International conference on machine learning}, pp.~201--210, PMLR, 2016.

\bibitem{zhao2021veriml}
L.~Zhao, Q.~Wang, C.~Wang, Q.~Li, C.~Shen, and B.~Feng, ``Veriml: Enabling integrity assurances and fair payments for machine learning as a service,'' {\em IEEE Transactions on Parallel and Distributed Systems}, vol.~32, no.~10, pp.~2524--2540, 2021.

\bibitem{liu2021zkcnn_zkpytorch_ref16}
T.~Liu, X.~Xie, and Y.~Zhang, ``{zkCNN: Zero Knowledge Proofs for Convolutional Neural Network Predictions and Accuracy},'' in {\em Proceedings of the 2021 ACM SIGSAC Conference on Computer and Communications Security (CCS '21)}, pp.~2968--2985, ACM, 2021.
\newblock Reference [16] in zkPyTorch.pdf, also cited as [17].

\bibitem{sun2024zkllm}
H.~Sun, J.~Li, and H.~Zhang, ``zkllm: Zero knowledge proofs for large language models,'' in {\em Proceedings of the 2024 on ACM SIGSAC Conference on Computer and Communications Security}, pp.~4405--4419, 2024.

\bibitem{lu2024efficient}
T.~Lu, H.~Wang, W.~Qu, Z.~Wang, J.~He, T.~Tao, W.~Chen, and J.~Zhang, ``An efficient and extensible zero-knowledge proof framework for neural networks,'' {\em Cryptology ePrint Archive}, 2024.

\bibitem{chen2024zkml}
B.-J. Chen, S.~Waiwitlikhit, I.~Stoica, and D.~Kang, ``Zkml: An optimizing system for ml inference in zero-knowledge proofs,'' in {\em Proceedings of the Nineteenth European Conference on Computer Systems}, pp.~560--574, 2024.

\bibitem{feng2021zen}
B.~Feng, L.~Qin, Z.~Zhang, Y.~Ding, and S.~Chu, ``Zen: Efficient zero-knowledge proofs for neural networks.,'' {\em IACR Cryptol. ePrint Arch.}, vol.~2021, p.~87, 2021.

\bibitem{feng2024zeno_zkpytorch_ref9}
B.~Feng, Z.~Wang, Y.~Wang, S.~Yang, and Y.~Ding, ``{ZENO: A Type-based Optimization Framework for Zero Knowledge Neural Network Inference},'' in {\em Proceedings of the 29th ACM International Conference on Architectural Support for Programming Languages and Operating Systems, Volume 1}, pp.~450--464, 2024.
\newblock Reference [9] in zkPyTorch.pdf.

\bibitem{hao2024scalable}
M.~Hao, H.~Chen, H.~Li, C.~Weng, Y.~Zhang, H.~Yang, and T.~Zhang, ``Scalable zero-knowledge proofs for non-linear functions in machine learning,'' in {\em 33rd USENIX Security Symposium (USENIX Security 24)}, pp.~3819--3836, 2024.

\bibitem{kang2022scaling}
D.~Kang, T.~Hashimoto, I.~Stoica, and Y.~Sun, ``Scaling up trustless dnn inference with zero-knowledge proofs,'' {\em arXiv preprint arXiv:2210.08674}, 2022.

\bibitem{zhang2020zero}
J.~Zhang, Z.~Fang, Y.~Zhang, and D.~Song, ``Zero knowledge proofs for decision tree predictions and accuracy,'' in {\em Proceedings of the 2020 ACM SIGSAC Conference on Computer and Communications Security}, pp.~2039--2053, 2020.

\bibitem{maheri2025telesparse}
M.~M. Maheri, H.~Haddadi, and A.~Davidson, ``Telesparse: Practical privacy-preserving verification of deep neural networks,'' {\em Proceedings on Privacy Enhancing Technologies (PETS)}, 2025.

\bibitem{abbaszadeh2024zero}
K.~Abbaszadeh, C.~Pappas, J.~Katz, and D.~Papadopoulos, ``Zero-knowledge proofs of training for deep neural networks,'' in {\em Proceedings of the 2024 on ACM SIGSAC Conference on Computer and Communications Security}, pp.~4316--4330, 2024.

\bibitem{garg2023experimenting}
S.~Garg, A.~Goel, S.~Jha, S.~Mahloujifar, M.~Mahmoody, G.-V. Policharla, and M.~Wang, ``Experimenting with zero-knowledge proofs of training,'' in {\em Proceedings of the 2023 ACM SIGSAC Conference on Computer and Communications Security}, pp.~1880--1894, 2023.

\bibitem{longpre2024data}
S.~Longpre, R.~Mahari, N.~Obeng-Marnu, W.~Brannon, T.~South, K.~Gero, S.~Pentland, and J.~Kabbara, ``Data authenticity, consent, \& provenance for ai are all broken: what will it take to fix them?,'' {\em arXiv preprint arXiv:2404.12691}, 2024.

\bibitem{stokes2021preventing}
J.~W. Stokes, P.~England, and K.~Kane, ``Preventing machine learning poisoning attacks using authentication and provenance,'' in {\em MILCOM 2021-2021 IEEE Military Communications Conference (MILCOM)}, pp.~181--188, IEEE, 2021.

\bibitem{gao2023retrieval}
Y.~Gao, Y.~Xiong, X.~Gao, K.~Jia, J.~Pan, Y.~Bi, Y.~Dai, J.~Sun, H.~Wang, and H.~Wang, ``Retrieval-augmented generation for large language models: A survey,'' {\em arXiv preprint arXiv:2312.10997}, vol.~2, no.~1, 2023.

\bibitem{gupta2024comprehensive}
S.~Gupta, R.~Ranjan, and S.~N. Singh, ``A comprehensive survey of retrieval-augmented generation (rag): Evolution, current landscape and future directions,'' {\em arXiv preprint arXiv:2410.12837}, 2024.

\bibitem{zhou2024trustworthiness}
Y.~Zhou, Y.~Liu, X.~Li, J.~Jin, H.~Qian, Z.~Liu, C.~Li, Z.~Dou, T.-Y. Ho, and P.~S. Yu, ``Trustworthiness in retrieval-augmented generation systems: A survey,'' {\em arXiv preprint arXiv:2409.10102}, 2024.

\bibitem{aws2024rag}
{Amazon Web Services}, ``What is rag? (retrieval-augmented generation),'' 2025.

\bibitem{aws2024ragauth}
{Amazon Web Services}, ``Authorizing access to data with rag implementations,'' 2025.

\bibitem{fitzgerald2024rag}
M.~Fitzgerald, ``The future of ai: Merging rag with regulatory standards \& compliance.'' LinkedIn, 2024.

\bibitem{kothapalli2024hypernova}
A.~Kothapalli and S.~Setty, ``Hypernova: Recursive arguments for customizable constraint systems,'' in {\em Annual International Cryptology Conference}, pp.~345--379, Springer, 2024.

\bibitem{kate2010constant}
A.~Kate, G.~M. Zaverucha, and I.~Goldberg, ``Constant-size commitments to polynomials and their applications,'' in {\em International conference on the theory and application of cryptology and information security}, pp.~177--194, Springer, 2010.

\bibitem{papamanthou2024reckle}
C.~Papamanthou, S.~Srinivasan, N.~Gailly, I.~Hishon-Rezaizadeh, A.~Salumets, and S.~Golemac, ``Reckle trees: Updatable merkle batch proofs with applications,'' in {\em Proceedings of the 2024 on ACM SIGSAC Conference on Computer and Communications Security}, pp.~1538--1551, 2024.

\bibitem{boneh2001short}
D.~Boneh, B.~Lynn, and H.~Shacham, ``Short signatures from the weil pairing,'' in {\em International conference on the theory and application of cryptology and information security}, pp.~514--532, Springer, 2001.

\bibitem{fiat1986prove}
A.~Fiat and A.~Shamir, ``How to prove yourself: Practical solutions to identification and signature problems,'' in {\em Conference on the theory and application of cryptographic techniques}, pp.~186--194, Springer, 1986.

\bibitem{schwartz1980fast}
J.~T. Schwartz, ``Fast probabilistic algorithms for verification of polynomial identities,'' {\em Journal of the ACM (JACM)}, vol.~27, no.~4, pp.~701--717, 1980.

\bibitem{jin2019pubmedqa}
Q.~Jin, B.~Dhingra, Z.~Liu, W.~W. Cohen, and X.~Lu, ``Pubmedqa: A dataset for biomedical research question answering,'' {\em arXiv preprint arXiv:1909.06146}, 2019.

\bibitem{grattafiori2024llama}
A.~Grattafiori, A.~Dubey, A.~Jauhri, A.~Pandey, A.~Kadian, A.~Al-Dahle, A.~Letman, A.~Mathur, A.~Schelten, A.~Vaughan, {\em et~al.}, ``The llama 3 herd of models,'' {\em arXiv preprint arXiv:2407.21783}, 2024.

\bibitem{hu2022lora}
E.~J. Hu, Y.~Shen, P.~Wallis, Z.~Allen-Zhu, Y.~Li, S.~Wang, L.~Wang, W.~Chen, {\em et~al.}, ``Lora: Low-rank adaptation of large language models.,'' {\em ICLR}, vol.~1, no.~2, p.~3, 2022.

\bibitem{kothapalli2022nova}
A.~Kothapalli, S.~Setty, and I.~Tzialla, ``Nova: Recursive zero-knowledge arguments from folding schemes,'' in {\em Annual International Cryptology Conference}, pp.~359--388, Springer, 2022.

\bibitem{nemecek2024topic}
A.~Nemecek, Y.~Jiang, and E.~Ayday, ``Topic-based watermarks for large language models,'' {\em arXiv preprint arXiv:2404.02138}, 2024.

\bibitem{nemecek2025feasibility}
A.~Nemecek, Y.~Jiang, and E.~Ayday, ``The feasibility of topic-based watermarking on academic peer reviews,'' {\em arXiv preprint arXiv:2505.21636}, 2025.

\bibitem{singh2025agentic}
A.~Singh, A.~Ehtesham, S.~Kumar, and T.~T. Khoei, ``Agentic retrieval-augmented generation: A survey on agentic rag,'' {\em arXiv preprint arXiv:2501.09136}, 2025.

\bibitem{lin-2004-rouge}
C.-Y. Lin, ``{ROUGE}: A package for automatic evaluation of summaries,'' in {\em Text Summarization Branches Out}, (Barcelona, Spain), pp.~74--81, Association for Computational Linguistics, July 2004.

\end{thebibliography}

\appendices

\section{Fine-Tuning Configuration}\label{appendixA}

This appendix provides the complete specifications and hyperparameters used during the fine-tuning process for both \texttt{Llama-3.2-1B} and \texttt{Llama-3.1-8B} models. All fine-tuning experiments were conducted on a single NVIDIA A100 GPU with 128GB memory. We employed Low-Rank Adaptation (LoRA) as our parameter-efficient fine-tuning method, targeting only the attention projection layers while keeping the base model frozen. Table~\ref{tab:hyperparameters} presents the comprehensive set of hyperparameters used throughout the training process. We utilized mixed-precision training to optimize memory usage and computational efficiency. Model checkpoints were saved at the end of each epoch, with the best model selected based on validation loss.

\begin{table}[h]
\centering
\caption{Fine-Tuning Hyperparameters}
\label{tab:hyperparameters}
\begin{tabular}{ll}
\hline
\textbf{Parameter} & \textbf{Value} \\
\hline
\multicolumn{2}{l}{\textit{General Training Parameters}} \\
Optimizer & AdamW \\
Learning Rate ($\eta$) & $5 \times 10^{-5}$\\
Number of Epochs ($E$) & 3 \\
Per-Device Batch Size & 2 \\
Gradient Accumulation Steps & 4 \\
Effective Batch Size ($B$) & 8 \\
Max Sequence Length & 2048 \\
Warmup Ratio & 0.2 \\
LR Scheduler Type & Linear \\
Mixed Precision & bfloat16 \\
Dataloader Workers & 4 \\
\hline
\multicolumn{2}{l}{\textit{LoRA Configuration}} \\
LoRA Rank ($r$) & 16 \\
LoRA Alpha ($\alpha$) & 32 \\
LoRA Dropout & 0.1 \\
Target Modules & q\_proj, k\_proj, v\_proj, o\_proj \\
Bias & None \\
Task Type & Causal LM \\
\hline
\multicolumn{2}{l}{\textit{Logging and Checkpointing}} \\
Evaluation Strategy & Epoch \\
Save Strategy & Epoch \\
Logging Steps & 10 \\
Save Total Limit & 2 \\
Metric for Best Model & Loss \\
Load Best Model at End & True \\
\hline
\end{tabular}
\end{table}

\section{Fine-Tuning Performance Validation}\label{appendixB}
To validate that fine-tuning improves model performance on the clinical domain while maintaining verifiable provenance, we evaluate both the base (non-fine-tuned) and fine-tuned variants on the downstream question-answering task using the same 100 held-out test samples employed in our cryptographic overhead measurements.

\subsection{Evalaution Methodology}
We assess model performance using ROUGE (Recall-Oriented Understudy for Gisting Evaluation)~\cite{lin-2004-rouge} scores, a standard metric for evaluating the quality of generated text against reference answers. ROUGE measures the overlap between generated responses and ground truth answers through n-gram matching: ROUGE-1 captures unigram overlap (individual word matches), ROUGE-2 measures bigram overlap (consecutive word pair matches), and ROUGE-L evaluates the longest common subsequence, capturing sentence-level structure similarity. These metrics are particularly appropriate for our biomedical question-answering task, as they quantify how well the model's responses align with expert-provided answers in terms of content coverage and structural coherence.

\subsection{Results}
Table~\ref{tab:rouge_metrics} presents the average ROUGE scores across all test samples for both model variants in their base and fine-tuned configurations. The results demonstrate consistent performance improvements from fine-tuning across all metrics. For 
\texttt{Llama-3.2-1B}, fine-tuning increased ROUGE-1 from 0.215 to 0.292, ROUGE-2 from 0.077 to 0.113, and ROUGE-L from 0.159 to 0.223. The \texttt{Llama-3.1-8B} model exhibited similar gains, with ROUGE-1 improving from 0.261 to 0.363, ROUGE-2 from 0.100 to 0.157, and ROUGE-L from 0.194 to 0.287.

\begin{table}[h]
\centering
\begin{tabular}{lccc}
\hline
\textbf{Model} & \textbf{ROUGE-1} & \textbf{ROUGE-2} & \textbf{ROUGE-L} \\
\hline
3.2-1B Base & 0.215 & 0.077 & 0.159 \\
3.2-1B Fine-tuned & \textbf{0.292} & \textbf{0.113} & \textbf{0.223} \\
3.1-8B Base & 0.261 & 0.100 & 0.194 \\
3.1-8B Fine-tuned & \textbf{0.363} & \textbf{0.157} & \textbf{0.287} \\
\hline
\end{tabular}
\caption{Average ROUGE scores for base and fine-tuned model variants. Bold values indicate best results.}
\label{tab:rouge_metrics}
\end{table}

\begin{figure}[t]
    \centering
    \begin{subfigure}[b]{0.48\textwidth}
        \centering
        \includegraphics[width=\textwidth]{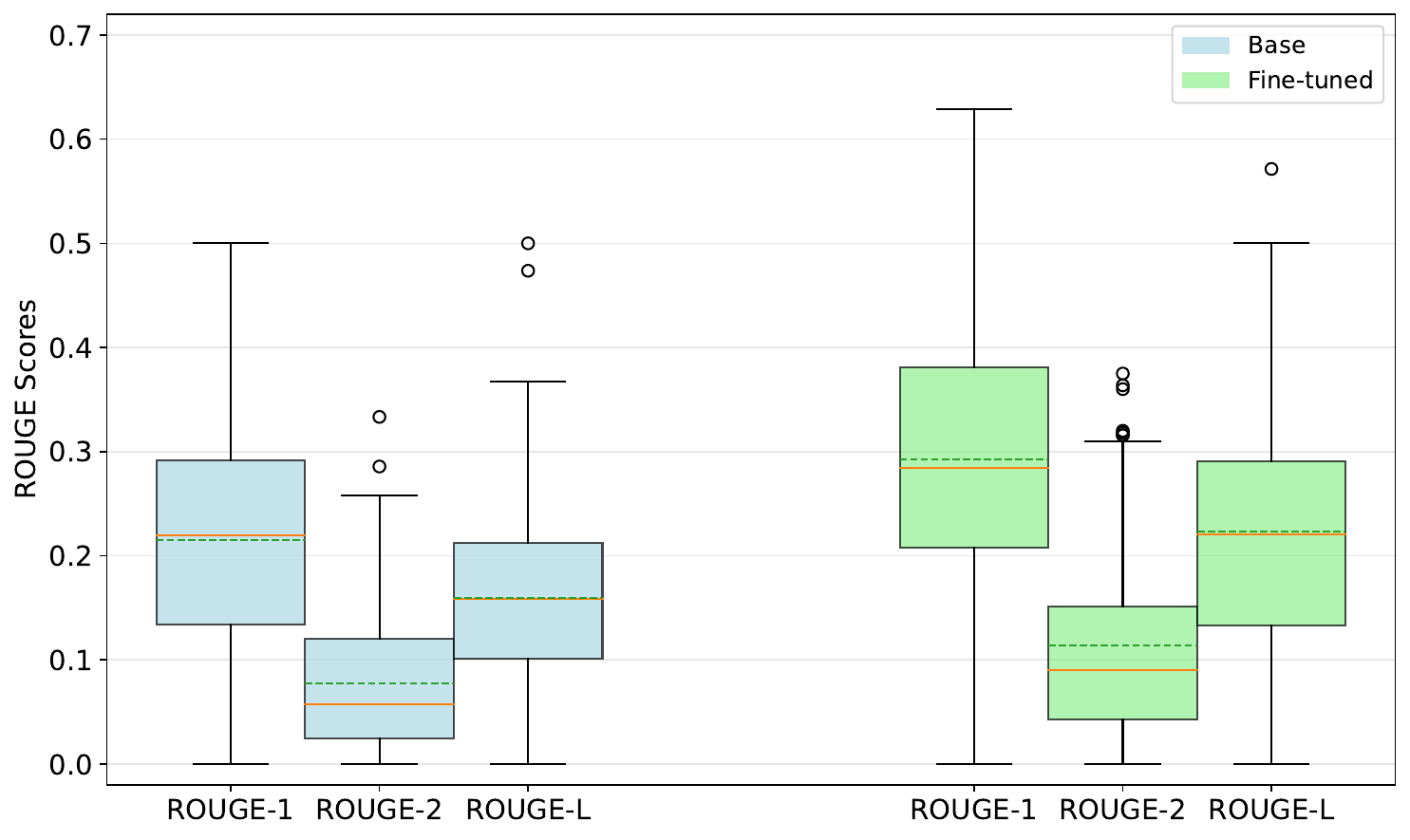}
        \caption{\texttt{Llama-3.2-1B}}
        \label{fig:rouge_boxplot_1b}
    \end{subfigure}
    \hfill
    \begin{subfigure}[b]{0.48\textwidth}
        \centering
        \includegraphics[width=\textwidth]{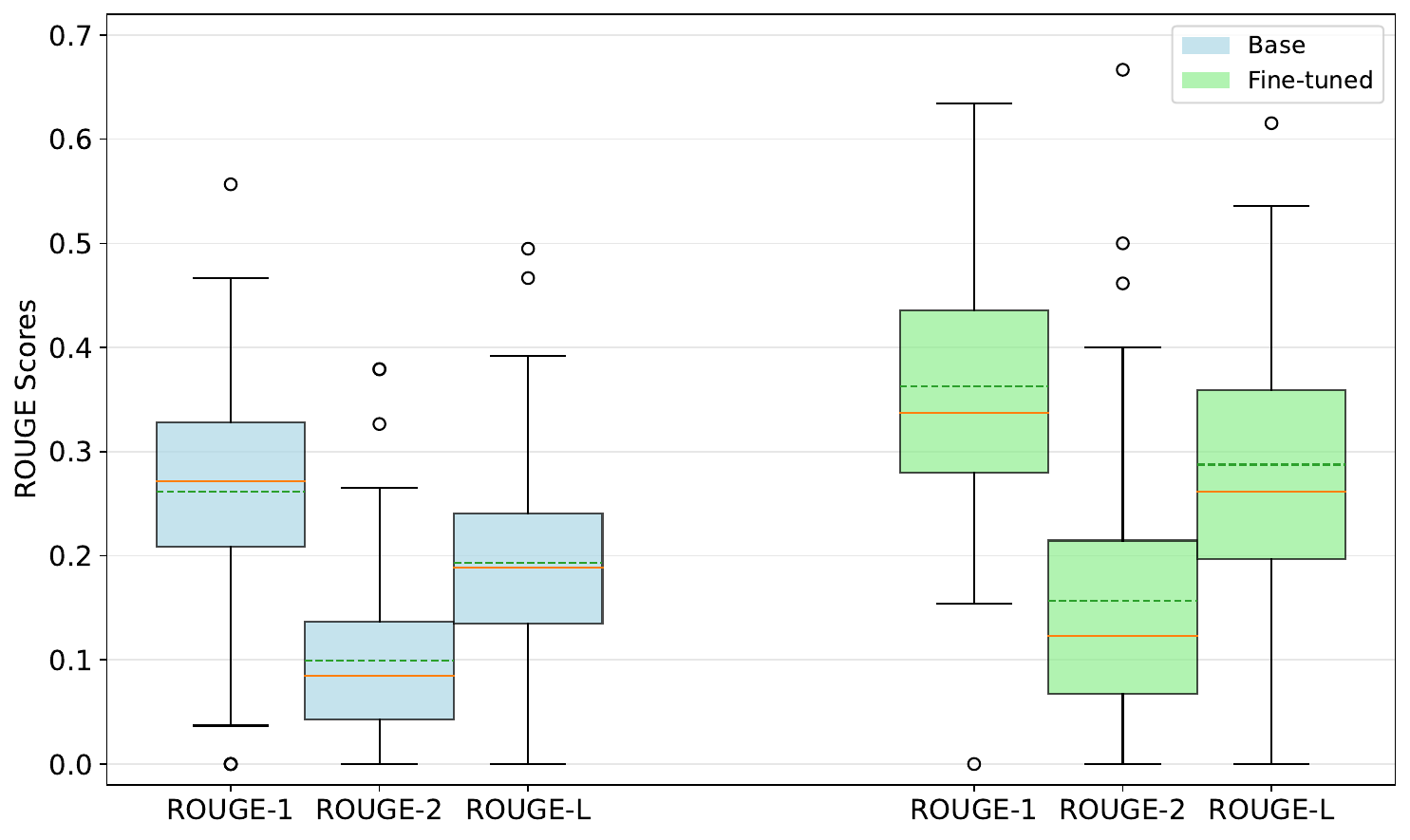}
        \caption{\texttt{Llama-3.1-8B}}
        \label{fig:rouge_boxplot_8b}
    \end{subfigure}
    \caption{Distribution of ROUGE scores across 100 test samples for base and fine-tuned model variants. Box plots show median (center line), interquartile range (box), and full range excluding outliers (whiskers) for ROUGE-1, ROUGE-2, and ROUGE-L metrics. Fine-tuning consistently improves performance across all metrics for both model sizes, with the \texttt{Llama-3.1-8B} variant achieving the highest scores and the most consistent performance, as evidenced by tighter distributions and elevated medians.}
    \label{fig:rouge_boxplots}
\end{figure}

Figure~\ref{fig:rouge_boxplots} illustrates the distribution of ROUGE scores across the test set for each model configuration. The box plots reveal that fine-tuning not only improves average performance but also reduces variance, producing more consistent response quality across diverse clinical queries. The fine-tuned \texttt{Llama-3.1-8B} model achieves the highest performance across all ROUGE metrics, demonstrating superior content alignment with expert answers. However, as shown in Table~\ref{tab:finetuned_comparison}, this performance advantage comes at a computational cost: the 8B model requires 1.499 seconds for inference compared to 0.820 seconds for the 1B model, representing an increase in inference time. When combined with proof-generation overhead (Table~\ref{tab:proof_combined_comparison}), this results in end-to-end times of 3.296 seconds versus 2.608 seconds for the largest sampling configuration, highlighting the inherent trade-off between model performance and computational efficiency in provenance-aware deployments.

The PubMedQA dataset additionally provides short-form structured answers (yes/no/maybe) alongside the long-form explanations. We evaluated exact match accuracy on these structured answers to assess the models' ability to produce precise clinical judgments. Base models for both architectures achieved 0\% accuracy on this task, failing to generate responses in the required format. The fine-tuned \texttt{lama-3.2-1B} model showed marginal improvement, but it still achieved 0\% accuracy. In contrast, the fine-tuned 
\texttt{Llama-3.1-8B} model achieved 43\% accuracy, demonstrating that sufficient model capacity combined with domain-specific fine-tuning enables the model to learn both the response format and domain-appropriate decision-making. This result further validates that fine-tuning on authenticated clinical datasets meaningfully improves model behavior on domain-specific tasks.

\section{Parameter Sampling Configuration}\label{appendixC}
We provide the exact parameter counts corresponding to each sampling percentage used in our experimental evaluation. As described in Section~\ref{parameter_sampling}, we sample weight differences from the query projection matrices (q\_proj) of the attention layers to evaluate ZKPROV's scalability while maintaining computational feasibility.

Table~\ref{tab:parameter_counts} presents the complete mapping between sampling percentages and the actual number of parameters sampled for both \texttt{Llama-3.2-1B} and \texttt{Llama-3.1-8B} models. The parameter counts reflect the total number of weight differences $\Delta W = W_{\text{fine-tuned}} - W_{\text{base}}$ extracted from the query projection matrices across all transformer layers. The \texttt{Llama-3.1-8B} model contains exactly $4\times$ the number of parameters as \texttt{Llama-3.2-1B} at each sampling percentage, reflecting the architectural differences between the two model variants.

\begin{table}[h]
\centering
\begin{tabular}{c|r|r|c}
\hline
\multirow{2}{*}{\textbf{Sampling \%}} & \multicolumn{2}{c|}{\textbf{Model}} & \multirow{2}{*}{\textbf{Proof Layers}} \\
\cline{2-3}
& \textbf{Llama-3.2-1B} & \textbf{Llama-3.1-8B} & \\
\hline
10\% & 419,430 & 1,677,721 & 8 \\
20\% & 838,860 & 3,355,443 & 16 \\
30\% & 1,258,291 & 5,033,164 & 24 \\
40\% & 1,677,721 & 6,710,886 & 32 \\
50\% & 2,097,152 & 8,388,608 & 40 \\
\hline
\end{tabular}
\caption{Parameter counts for different sampling percentages across model variants. Each row shows the exact number of parameters sampled from query projection matrices and the corresponding number of proof layers used in the recursive SNARK construction.}
\label{tab:parameter_counts}
\end{table}

These parameter counts directly support our claims of sublinear scaling in proof generation time (Table~\ref{tab:proof_combined_comparison}) and model size independence in cryptographic overhead. Despite the $4\times$ difference in sampled parameters between the two model variants at equivalent sampling percentages, proof generation times remain nearly identical, demonstrating that the cryptographic complexity depends primarily on the proof structure rather than the absolute parameter count.

\end{document}